\documentclass[a4paper,fleqn,usenatbib,useAMS]{mnras}
\usepackage[T1]{fontenc}
\usepackage{ae,aecompl}
\usepackage{graphicx}
\usepackage{amsmath}
\usepackage{amssymb}
\usepackage{lineno}

\newcommand\xmm{{\it XMM-Newton}}

\newcommand\s{{\rm~s}}
\newcommand\ks{{\rm~ks}}

\newcommand\hz{{\rm~Hz}}

\newcommand\keV{{\rm~keV}}

\title[X-ray lags in low-mass AGNs]{Discovery of soft and hard X-ray time lags in low-mass AGNs}

\author[L. Mallick et al.]{L. Mallick$^{1,2}$,\thanks{Email: labani.mallick@iiap.res.in (LM)}
	D. R. Wilkins$^{3}$,
	W. N. Alston$^{4,5}$,
  A. Markowitz$^{6,7}$,
  B. De Marco$^{6,8}$,
\newauthor  M. L. Parker$^{5,4}$,
  A. M. Lohfink$^{9}$,
  and C. S. Stalin$^{1}$ \\
$^{1}$ Indian Institute of Astrophysics, Block II, Koramangala, Bengaluru 560034, India\\
$^{2}$ Department of Astronomy and Astrophysics, Pennsylvania State University, 525 Davey Laboratory, University Park, PA 16802, USA\\
$^{3}$ Kavli Institute for Particle Astrophysics and Cosmology, Stanford University, 452 Lomita Mall, Stanford, CA 94305, USA\\
$^{4}$ European Space Agency (ESA), European Space Astronomy Centre (ESAC), E-28691 Villanueva de la Ca\~{n}ada, Madrid, Spain\\
$^{5}$ Institute of Astronomy, University of Cambridge, Madingley Road, Cambridge CB3 0HA, UK\\
$^{6}$ Nicolaus Copernicus Astronomical Center, Polish Academy of Sciences, Bartycka 18, 00-716, Warsaw, Poland \\
$^{7}$ University of California, San Diego, Center for Astrophysics and Space Sciences, 9500 Gilman Dr, La Jolla, CA 92093, USA \\
$^{8}$ Departament de F\'isica, EEBE, Universitat Polit\'ecnica de Catalunya, Av. Eduard Maristany 16, E-08019, Barcelona, Spain \\
$^{9}$ eXtreme Gravity Institute, Department of Physics, Montana State University, Bozeman, MT 59717, USA 
}

\begin{document}

\date{Accepted 2021 March 1. Received 2021 February 26; in original form 2020 October 29}


\pagerange{\pageref{firstpage}--\pageref{lastpage}} 

\maketitle



\label{firstpage}

\begin{abstract}

The scaling relations between the black hole (BH) mass and soft lag properties for both active galactic nuclei (AGNs) and BH X-ray binaries (BHXRBs) suggest the same underlying physical mechanism at work in accreting BH systems spanning a broad range of mass. However, the low-mass end of AGNs has never been explored in detail. In this work, we extend the existing scaling relations to lower-mass AGNs, which serve as anchors between the normal-mass AGNs and BHXRBs. For this purpose, we construct a sample of low-mass AGNs ($M_{\rm BH}<3\times 10^{6} M_{\rm \odot}$) from the \xmm{} archive and measure frequency-resolved time delays between the soft (0.3$-$1\keV{}) and hard (1$-$4\keV{}) X-ray emissions. We report that the soft band lags behind the hard band emission at high frequencies $\sim[1.3-2.6]\times 10^{-3}$\hz{}, which is interpreted as a sign of reverberation from the inner accretion disc in response to the direct coronal emission. At low frequencies ($\sim[3-8]\times 10^{-4}$\hz{}), the hard band lags behind the soft band variations, which we explain in the context of the inward propagation of luminosity fluctuations through the corona. Assuming a lamppost geometry for the corona, we find that the X-ray source of the sample extends at an average height and radius of $\sim 10r_{\rm g}$ and $\sim 6r_{\rm g}$, respectively. Our results confirm that the scaling relations between the BH mass and soft lag amplitude/frequency derived for higher-mass AGNs can safely extrapolate to lower-mass AGNs, and the accretion process is indeed independent of the BH mass.

\end{abstract}


\section{Introduction}
Active galactic nuclei (AGNs) emit radiation across the entire electromagnetic spectrum, and they are highly variable in the X-ray waveband. The observed fast variability indicates that the X-ray emitting region is compact and located in the vicinity of the central supermassive black hole (SMBH). The measurement of time delays between different X-ray energy bands as a function of Fourier frequency can shed light not only on the physical processes at work in the strong gravitational field of the BH but also on the geometry of the accretion disc-corona system (e.g., \citealt{wf13,em14,wg15,al20}). When the hard band (1$-$4\keV{}) dominated by primary coronal emission, lags behind the soft band (0.3$-$1\keV{}), we observe a positive or hard lag. Hard lags have been observed both in stellar-mass X-ray binaries (e.g., \citealt{mi88,no99}) and AGNs (e.g., \citealt{pa01,mc04}). The origin of hard lags has been attributed to the standard Comptonization process within the X-ray corona \citep{no99} or mass accretion rate fluctuations propagating inwards through the accretion disc \citep{ko01,au06,hogg16}. On the other hand, we observe a soft or negative lag when the variations in the reflection-dominated soft band lag behind the variations in the primary emission-dominated hard band. These have now been observed in many AGNs (\citealt{fa09,dm13,ka16,wi17,ma18}) as well as in stellar-mass BH binaries \citep{ut11,dm15}. The soft lag can be explained as a signature of reverberation of X-ray photons from the accretion disc (e.g., \citealt{zo10,ca13,dm15}) and/or as thermally reprocessed emission from the warm corona \citep{gd14}. An alternative interpretation of the soft lag is scattering of X-rays from a distant ($10-100$ gravitational radii), partially-covering absorbing medium \citep{mi10}. However, this model is unable to explain the fundamental properties of the observed variability process: the linear root mean squared (rms)--flux relation and the log-normal flux distribution \citep{ut01,ut05,al19}.

The unification theory of BH accretion suggests that the accretion process is independent of the BH mass, and we expect to observe similar timing properties in sources spanning a wide range of BH mass (e.g., \citealt{mc06}). The discovery of scaling relations between the BH mass and soft lag properties for both supermassive (\citealt{dm13}; hereafter DM13) and stellar-mass BHs \citep{dm15} is a breakthrough in this context. However, it is still unknown how the lower-mass SMBHs fit into these scaling relations. The lower-mass AGNs are intriguing sources that demand our attention for several reasons. They have great potential to provide crucial constraints on the nature of primordial SMBHs for models of cosmological BH growth. They can serve as anchors to test if the scaling relations derived for higher-mass AGNs can safely extrapolate to stellar-mass BH binaries and validate the existence of prevalent accretion properties at all mass scales. For a given radius, the inner disc temperature of lower-mass AGNs is higher compared to higher-mass AGN discs, which could potentially impact coronal geometry and/or reflection emission properties of lower-mass SMBHs. We thus aim to extend the existing BH mass vs. soft lag scaling relations of DM13 to lower-mass SMBHs ($\log M_{\rm BH}\sim 5-6$) and obtain a complete assessment of the lag-mass correlations for a broader mass range ($\log M_{\rm BH}\sim 5-8$). In this paper, we present the analysis of lag-frequency spectra and report the discovery of both soft and hard lags in several low-mass AGNs. 

We organize the paper as follows: In Section~\ref{sec2}, we describe the sample selection and data reduction method. In Section~\ref{sec3}, we describe lag-frequency spectral analysis. We present and discuss our results in Section~\ref{sec4}. We conclude and summarize the main findings in Section~\ref{sec5}.

\begin{table*}
\centering
\caption{Source sample employed in this work. Columns~(1) and (2) show the source name and total filtered exposure length in ks, respectively. Columns~(3), (4) and (5) show the background-subtracted EPIC pn$+$MOS counts in the full (0.3$-$10\keV{}), soft (0.3$-$1\keV{}) and hard (1$-$4\keV{}) bands, respectively. Columns~(6), (7) and (8) show the optical classification obtained from NED/SIMBAD, source redshift and Eddington ratio, respectively. Column~(9) shows the central black hole mass. We take the Eddington ratio and BH mass for POX52 from \citet{ba04}; for all other objects from the GH07 catalog. The scatter in the BH mass is $\sim 0.3$~dex for the GH07 sample.}
\label{tab1}           
\scalebox{0.95}{%
\begin{tabular}{c c c c c c c c c}
\hline 
Name & Total filtered exp. (ks) & Total counts & Soft counts & Hard counts & Type & Redshift & $L_{\rm bol}/L_{\rm E}$ & $M_{\rm BH}$ ($10^{5}M_{\rm \odot}$)\\  
(1) & (2) & (3) & (4) & (5) & (6) & (7) & (8) & (9)\\
\hline 
J0107 & 34.8 & $1.6\times10^4$ & $9.5\times10^3$ & $5.6\times10^3$ & NLSy~1 & 0.0767 & 0.40 & 15.8$^{+15.7}_{-7.9}$\\ [0.2cm]

J0942 & 55.0 & $7.6\times10^3$ & $4.7\times10^3$ & $2.4\times10^3$ & Sy~1 & 0.197 & 0.63 & 15.8$^{+15.7}_{-7.9}$\\ [0.2cm]

J1023 & 219.4 & $3.1\times10^4$ & $2.0\times10^4$ & $1.0\times10^4$ & NLSy~1 & 0.0989 & 0.50 & 5.0$^{+4.9}_{-2.5}$\\ [0.2cm]

J1140 & 150.7 & $1.1\times10^5$ & $7.8\times10^4$ & $3.1\times10^4$ & NLSy~1 & 0.081 & 0.63 & 12.6$^{+12.5}_{-6.3}$\\ [0.2cm]

J1347 & 30.8 & $2.3\times10^4$ & $1.5\times10^4$ & $7\times10^3$ & NLSy~1 & 0.0643 & 0.50 & 10.0$^{+9.9}_{-5.0}$\\ [0.2cm]

J1434 & 47.9 & $8.5\times10^3$ & $4.4\times10^3$ & $3.5\times10^3$ & NLSy~1 & 0.0283 & 0.10 & 6.3$^{+6.2}_{-3.1}$\\ [0.2cm]

J1559 & 208.6 & $7.9\times10^5$ & $5.0\times10^5$ & $2.6\times10^5$ & NLSy~1 & 0.031 & 0.63 & 15.8$^{+15.7}_{-7.9}$\\ [0.2cm]

POX52 & 85.3 & $7.6\times10^3$ & $2.8\times10^3$ & $3.2\times10^3$ & Sy~1.8 & 0.021 & 0.75 & 1.6$^{+1.1}_{-1.1}$\\ [0.2cm]
\hline 
\end{tabular}}
\end{table*}

\begin{table*}
\centering
\begin{center}
\caption{Columns (2) and (3) show detected soft ($\nu_{\rm s}$) and hard ($\nu_{\rm h}$) lag frequencies in mHz, respectively. Columns (4) and (5) show the measured soft and hard lag amplitudes in seconds, respectively. The 1$\sigma$ uncertainty in lags was obtained through Monte Carlo simulations. The direct-to-reflected flux fraction ($\delta_{\rm soft}$) in the soft band and reflected-to-direct flux fraction ($\delta_{\rm hard}$) in the hard band are quoted in columns (6) and (7), respectively. Columns (8) and (9) show the coronal height ($h_{\rm c}$) and radius ($r_{\rm c}$) in units of $r_{\rm g}$, respectively. Columns (10) and (11) show the confidence intervals of detected soft and hard lags, respectively.}
\label{tab2}           
\scalebox{0.95}{%
\begin{tabular}{ccccccccccc}
\hline 
Source  & $\nu_{\rm s}$ (mHz) & $\nu_{\rm h}$ (mHz) & $|\tau_{\rm soft}|$ (s) & $\tau_{\rm hard}$ (s) & $\delta_{\rm soft}$  &  $\delta_{\rm hard}$ & $h_{\rm c}$ ($r_{\rm g}$) & $r_{\rm c}$ ($r_{\rm g}$) & C.I.-soft  & C.I.-hard \\ 
  &   &  &  & & & &  & & (\%) & (\%) \\ 
(1)    & (2)  & (3) & (4) & (5) & (6) & (7) & (8) & (9) & (10) & (11) \\
\hline 
J0107 & 1.55$\pm 0.49$ & 0.57$\pm 0.49$ & 65.2$\pm 46.7$  & 72.5$\pm 50.6$   & 0.3  & 0.6   & 9.2$\pm 6.6$  & 4.6$\pm 2.5$   &  91.5 & 94.6  \\ [0.2cm]

J0942 & 1.28$\pm 0.41$ & 0.45$\pm 0.41$ & $<84.6$    & 289.3$\pm 75.8$      &  $-$ & $-$  &    $-$         & $-$           &  64.0 & 99.9  \\ [0.2cm]

J1023 & 2.65$\pm 0.47$ & 0.78$\pm 0.47$ & 74.5$\pm 23.6$  & $<113.9$        & 0.62  & 0.76  & 23.2$\pm 7.4$ & $<14.5$      &  99.6  & 56.0  \\ [0.2cm]

J1140 & 1.81$\pm 0.35$ & 0.39$\pm 0.35$ & 35.2$\pm 19.9$  & 133.6$\pm 40.0$ &  0.79 & 0.69  & 2.3$\pm 1.3$   & 2.6$\pm 1.3$  &  92.6 & 99.9   \\ [0.2cm]

J1347 & 1.55$\pm 0.49$ & 0.57$\pm 0.49$ & 74.0$\pm 34.6$  & 74.5$\pm 36.4$ &  0.63 &  0.14  & 9.9$\pm 4.6$  & 7.5$\pm 3.9$  &  98.1 & 99.2   \\ [0.2cm]

J1434 & 2.58$\pm 0.48$ & 0.64$\pm 0.48$ & 51.6$\pm 36.0$  & $<136.3$       & 0.72 &  0.1    & 10.1$\pm 7.0$ &  $< 16.4$    &  90.9 & 66.1  \\ [0.2cm]

J1559 & 2.21$\pm 0.31$ & 0.35$\pm 0.31$ & 22.1$\pm 9.7$  & 80.7$\pm 22.5$  & 0.30  & 0.43   & 2.2$\pm 1.0$  & 2.2$\pm 1.1$  &  97.5 & 99.2   \\ [0.2cm]

POX52 & 1.55$\pm 0.49$ &  $-$           & 50.4$\pm 35.2$ & $-$             &  $-$ & $-$  & $-$  & $-$   &  88.1 &  $-$   \\ [0.2cm]
\hline 
\end{tabular}}
\end{center} 
\end{table*}

\begin{table}
\centering
\begin{center}
\caption{Parameters as obtained from the fitting of the lag profile with a monotonically declining hard lag model, $\tau_{h}=k(\nu/\nu_{0})^{-\alpha}$, indicating poor fits in all cases.}
\label{tab3}           
\scalebox{0.99}{%
\begin{tabular}{cccccc}
\hline 
Source  & Norm ($k$) & $\nu_{0}$ (Hz) & Slope ($\alpha$) & $\chi^{2}$/d.o.f \\ 
\hline 
J0107 & $ 4.8$ & $2\times 10^{-3}$ & $ 2.19$  & 20.2/2     \\ [0.15cm]

J0942 & $ 13.4$ & $ 1.9\times 10^{-3}$ & $ 2.19$ &  22.8/3    \\ [0.15cm]

J1023 & $ 0.53$ & $ 2\times 10^{-3}$ & $ 4.9$  & 27.0/2   \\ [0.15cm]

J1140 & $ 2.0$ & $ 1.9\times 10^{-3}$  & $ 2.67$  & 8.9/4  \\ [0.15cm]

J1347 & $ 37.2$ & $ 1.3\times 10^{-3}$ & $ 0.77$  & 87.1/2  \\ [0.15cm]

J1434 & $ 0.46$ & $ 1.6\times 10^{-3}$ & $ 4.9$  & 26.1/2     \\ [0.15cm]

J1559 & $ 5.9$ & $ 1.2\times 10^{-3}$ & $ 2.14$  & 16.8/5   \\ [0.15cm]

POX52 & $ 1.0$ & $ 1.7\times 10^{-3}$ & $ -1.11$  & 12.0/2    \\ [0.15cm]
\hline 
\end{tabular}}
\end{center} 
\end{table}

\section{Data Sample and reduction}
\label{sec2}
We selected a sample of very low-mass AGNs ($M_{\rm BH}<3\times 10^{6} M_{\rm \odot}$) from the Sloan Digital Sky Survey (SDSS) that were cataloged by \citet{gh07} (hereafter GH07) and one AGN (POX~52) discovered in the POX objective-prism survey \citep{ksk81}. A thorough search of the \xmm{} \citep{ja01} archive resulted in a sample of 26 objects having one or multiple observations. The \xmm{}/European Photon Imaging Camera (EPIC) data for the 26 objects were reduced with the Scientific Analysis System ({\tt{SAS}}~v.18.0.0) and the updated (as of 2020 May 12) calibration files. We processed the data obtained from both EPIC pn and MOS with the {\tt{SAS}} tasks {\tt{epproc}} and {\tt{emproc}}, respectively. We rejected the bad pixel events by setting {\tt{FLAG}}$==$0, and filtered the processed pn and MOS events using {\tt{PATTERN}}$\leq$4 and {\tt{PATTERN}}$\leq$12, respectively. To search for the flare-corrected good time intervals ({\tt{GTI}}), we adopted a 3$\sigma$ clipping method which can effectively eliminate the high-count-rate tail of the flare histogram while retaining useful events \citep{ctc18}. We first created single-pixel event light curves for the 10$-$12\keV{} band with time bins of 100\s{} and rejected those time intervals whose 10$-$12\keV{} count rates exceeded 3$\sigma$ above the mean. We then applied the 3$\sigma$ clipped {\tt{GTI}} to the 0.3$-$10\keV{} light curves and obtained the flare-filtered cleaned event files. This criterion excluded three sources massively dominated by background flares and led to a sample consisting of 23 AGNs. The source and background area for each observation were selected from a circular region of radius 20~arcsec centered on the point source and nearby source-free zone, respectively. We extracted the deadtime-corrected, background-subtracted source light curves with time bin size of 100\s{} using the {\tt{SAS}} task {\tt{epiclccorr}}. The EPIC-pn and MOS light curves were then added using the {\tt{lcmath}} tool to increase the signal-to-noise. The resulting light curves contain periods when both the detectors were functioning. A small number of data gaps were replaced by the method of linear interpolation implemented in the X-ray timing analysis package {\tt{PYLAG}}\footnote[1]{\url{http://github.com/wilkinsdr/pylag}}. We excluded Seyfert~2 AGNs from further analysis to concentrate only on sources with unobscured views of their central engines. To probe variability power above the Poisson noise level of 20\hz{}$^{-1}$, we applied a stringent total filtered exposure threshold of 30\ks{} and ensure that the background-subtracted average source counts in the 0.3$-$10\keV{} band is at least $3\times 10^3$. These criteria lead to a final sample of 8 AGNs which are Seyfert~1 galaxies. The characteristics and observational details of the final 8 sources are listed in Table~\ref{tab1} and Table~\ref{tabA1}, respectively. For the GH07 sample, the central BH masses were estimated (see equation~A1 of GH07) using the $L_{5100\textrm{\AA}}$--$L_{\alpha}$ relation from \citet{gh05} in combination with the revised $R_{\rm BLR}$-$L_{\alpha}$ relation from \citet{be06}, where $L_{5100}\textrm{\AA}$ is the nonstellar continuum luminosity at 5100$\textrm{\AA}$, $R_{\rm BLR}$ is the broad line region (BLR) size and $L_{\alpha}$ is the $H_{\alpha}$ line luminosity. We adopted a typical BH mass uncertainty of 0.3~dex \citep{gh06} incurred by the BLR geometrical factor in the $R_{\rm BLR}$-$L_{\alpha}$ relation. The BH mass of POX~52 was derived from $R_{\rm BLR}$ and the $H_{\beta}$ linewidth (equation~2 of \citealt{ba04}) where $R_{\rm BLR}$ was inferred using the calibrated $L_{5100\textrm{\AA}}$ versus $R_{\rm BLR}$ relation from \citet{ka00}. The BH mass uncertaity for POX~52 was obtained from the fitting of updated $M_{\rm BH}-\sigma_{\star}$ relation \citep{tr02} where both the measurement error on the host galaxy velocity dispersion ($\sigma_{\star}$) for POX~52 and the error in the fitted parameters of the $M_{\rm BH}-\sigma_{\star}$ relation contribute to the BH mass uncertainty.

\begin{figure*}
\centering
\begin{center}
\includegraphics[scale=0.32,angle=-0]{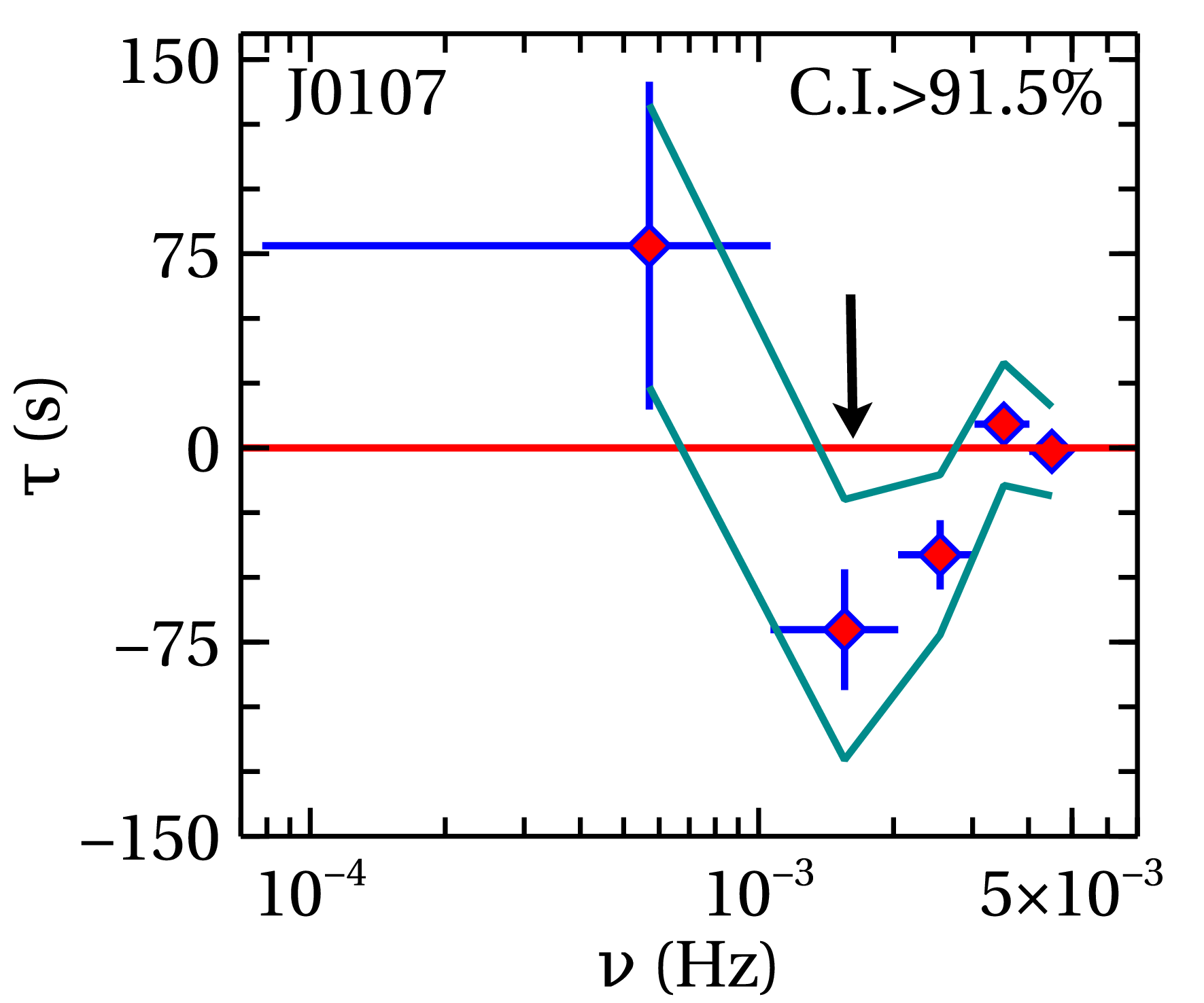}
\includegraphics[scale=0.32,angle=-0]{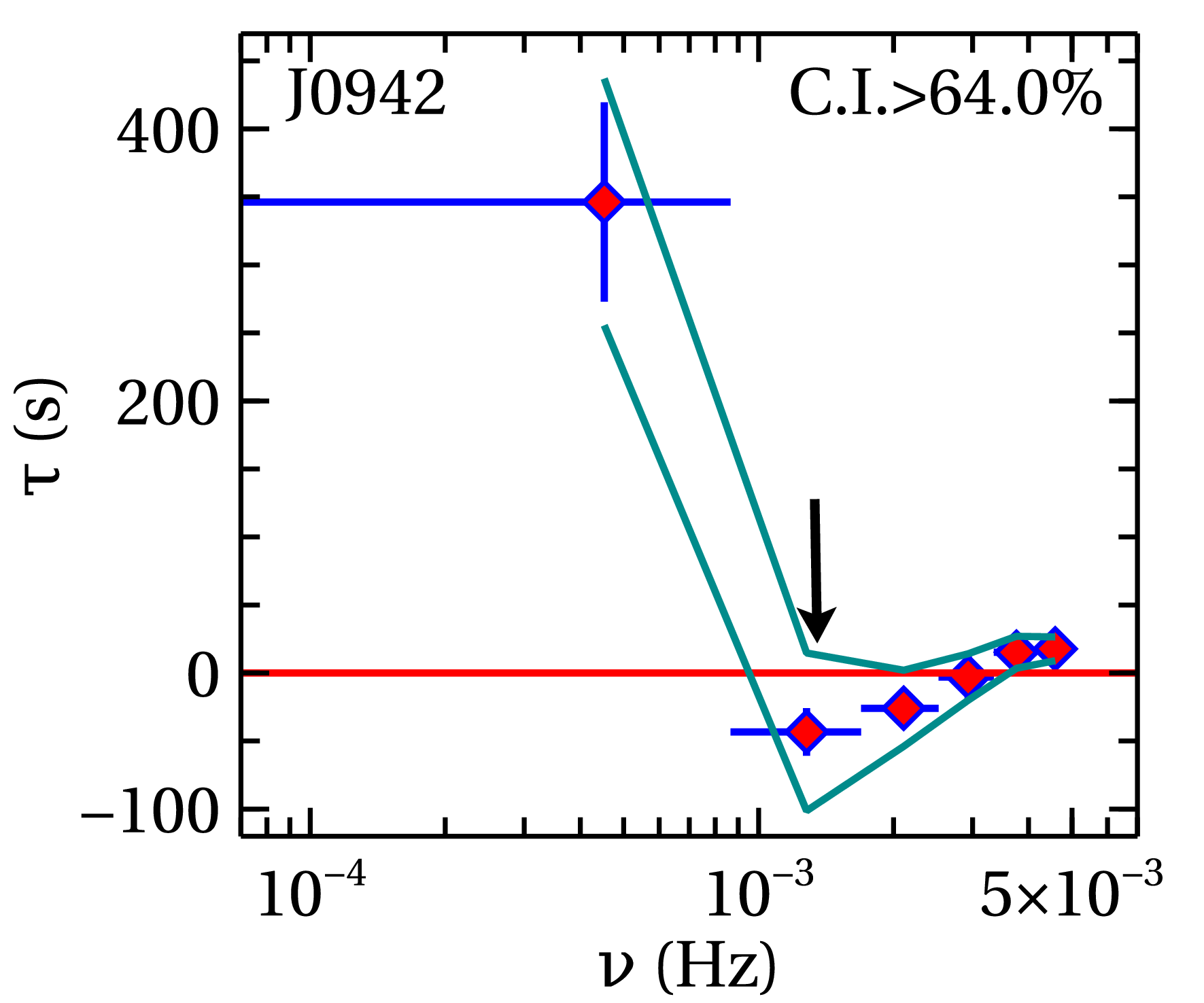}
\includegraphics[scale=0.32,angle=-0]{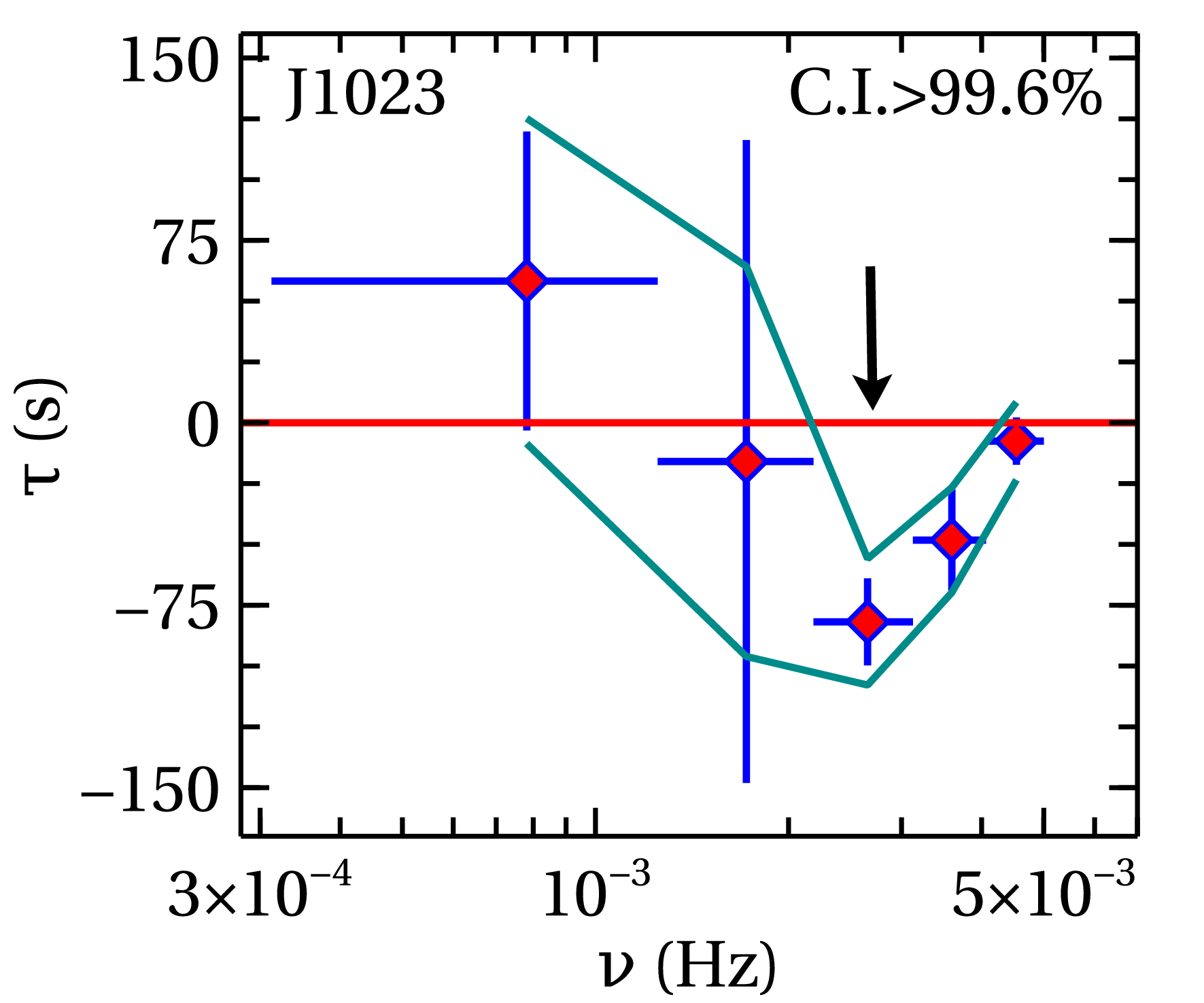}
\includegraphics[scale=0.32,angle=-0]{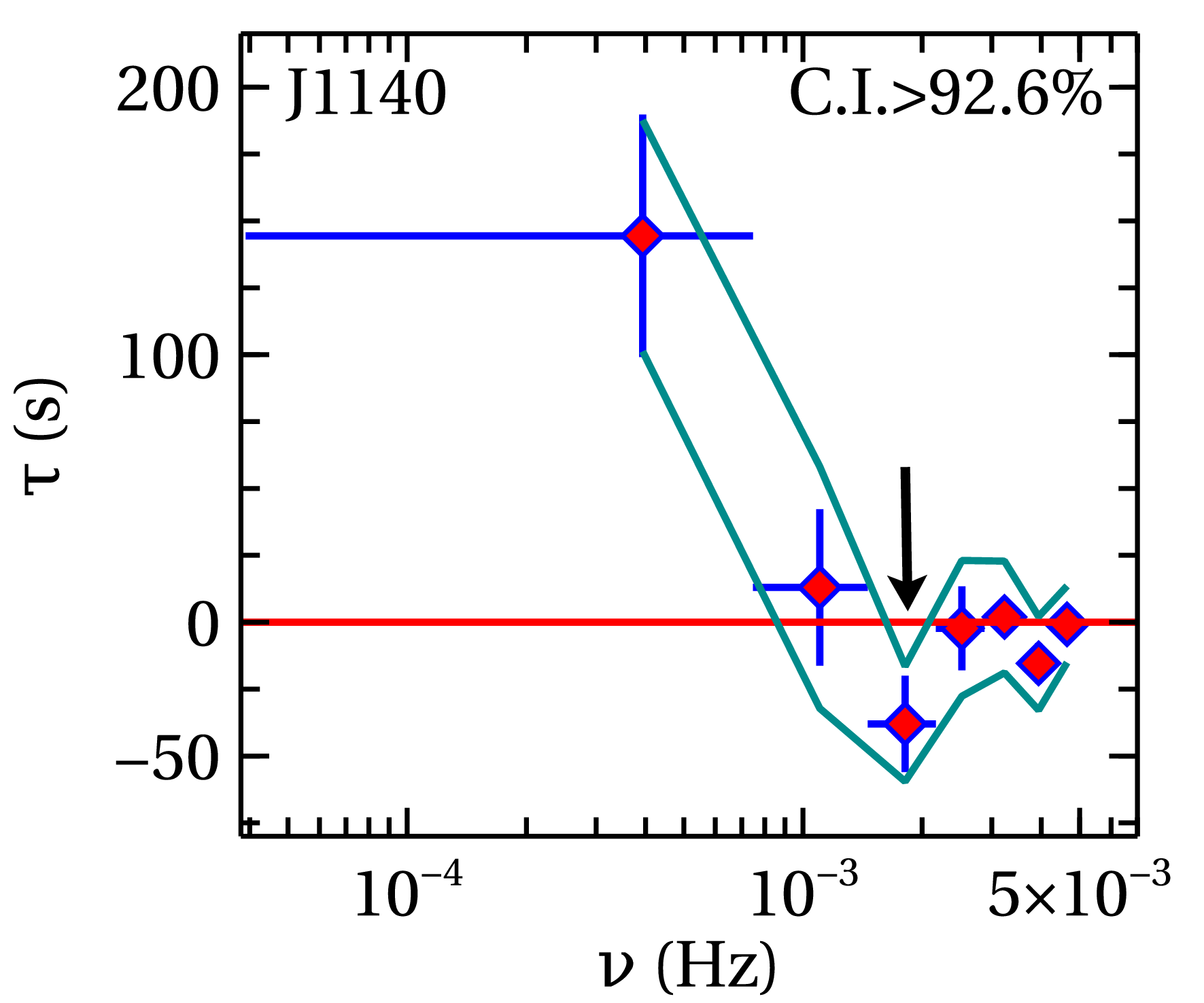}
\includegraphics[scale=0.32,angle=-0]{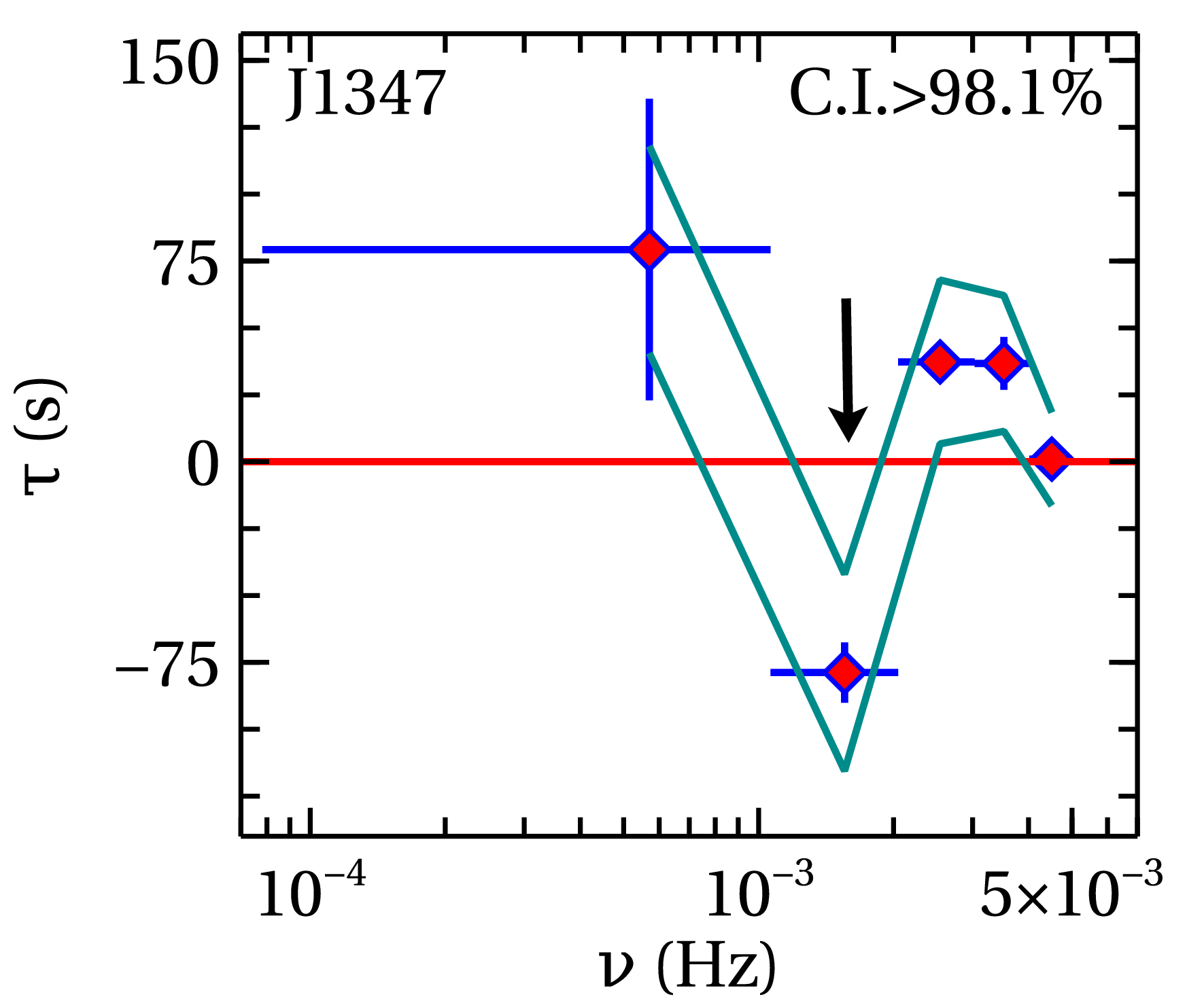}
\includegraphics[scale=0.32,angle=-0]{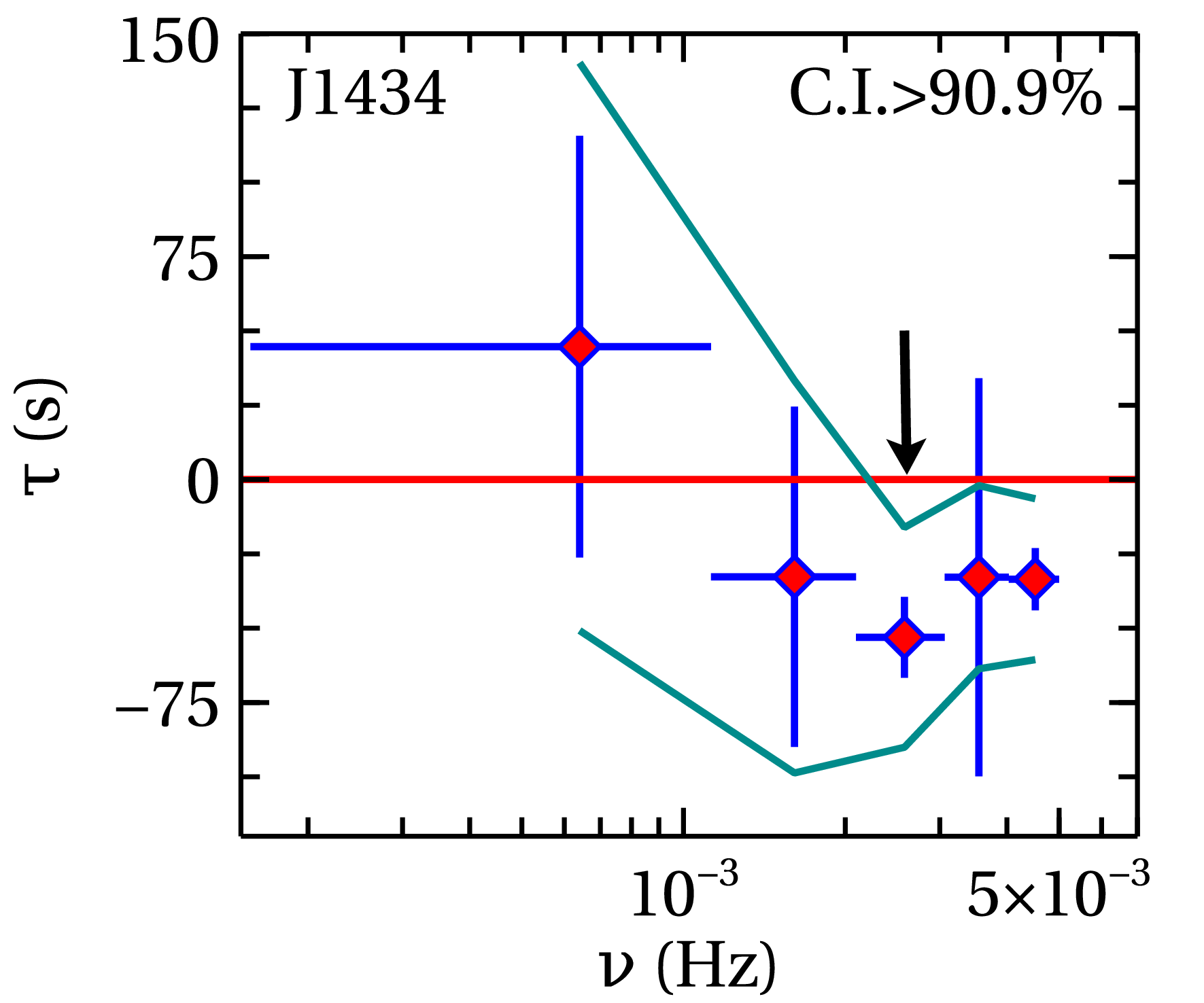}
\includegraphics[scale=0.32,angle=-0]{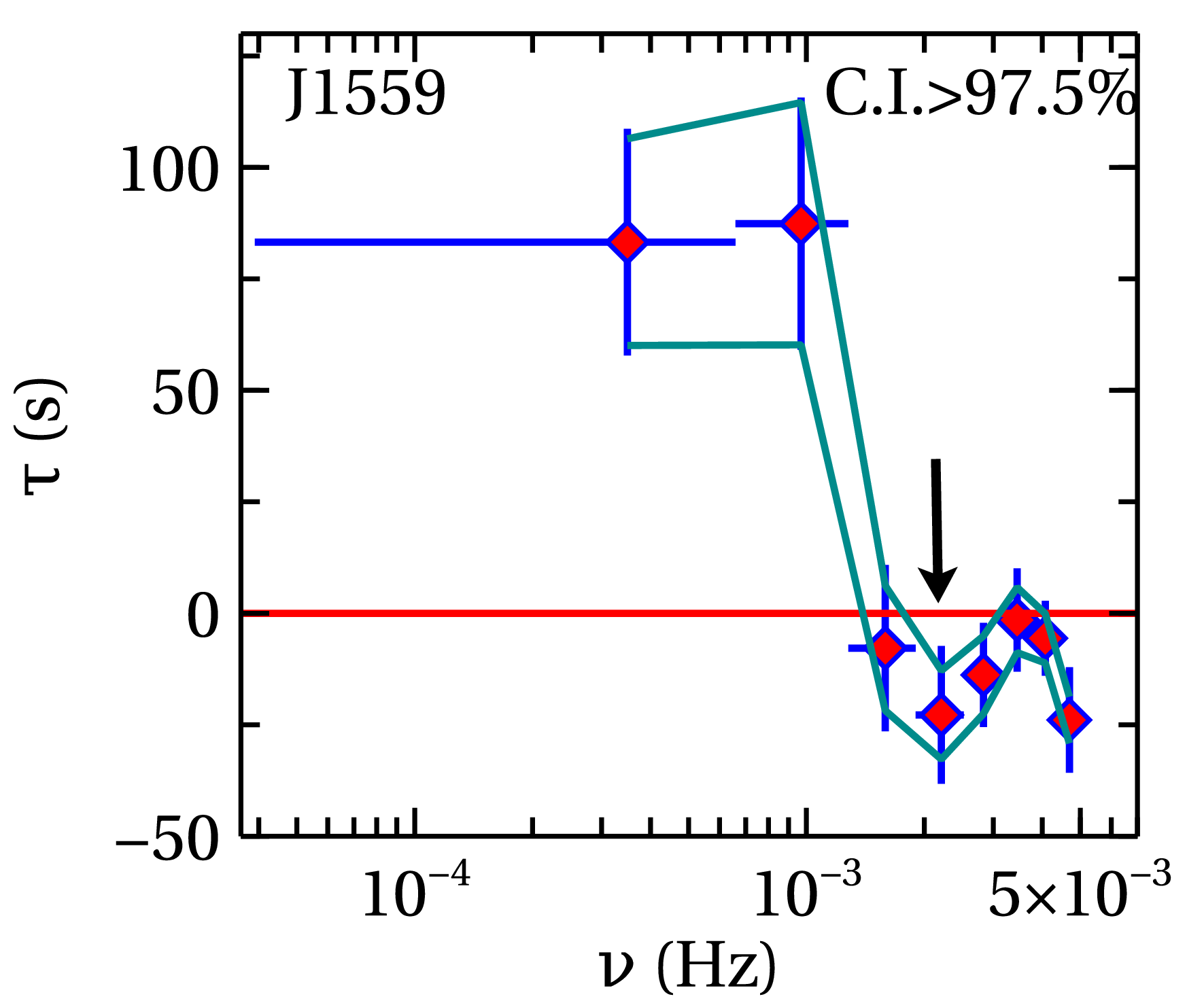}
\includegraphics[scale=0.32,angle=-0]{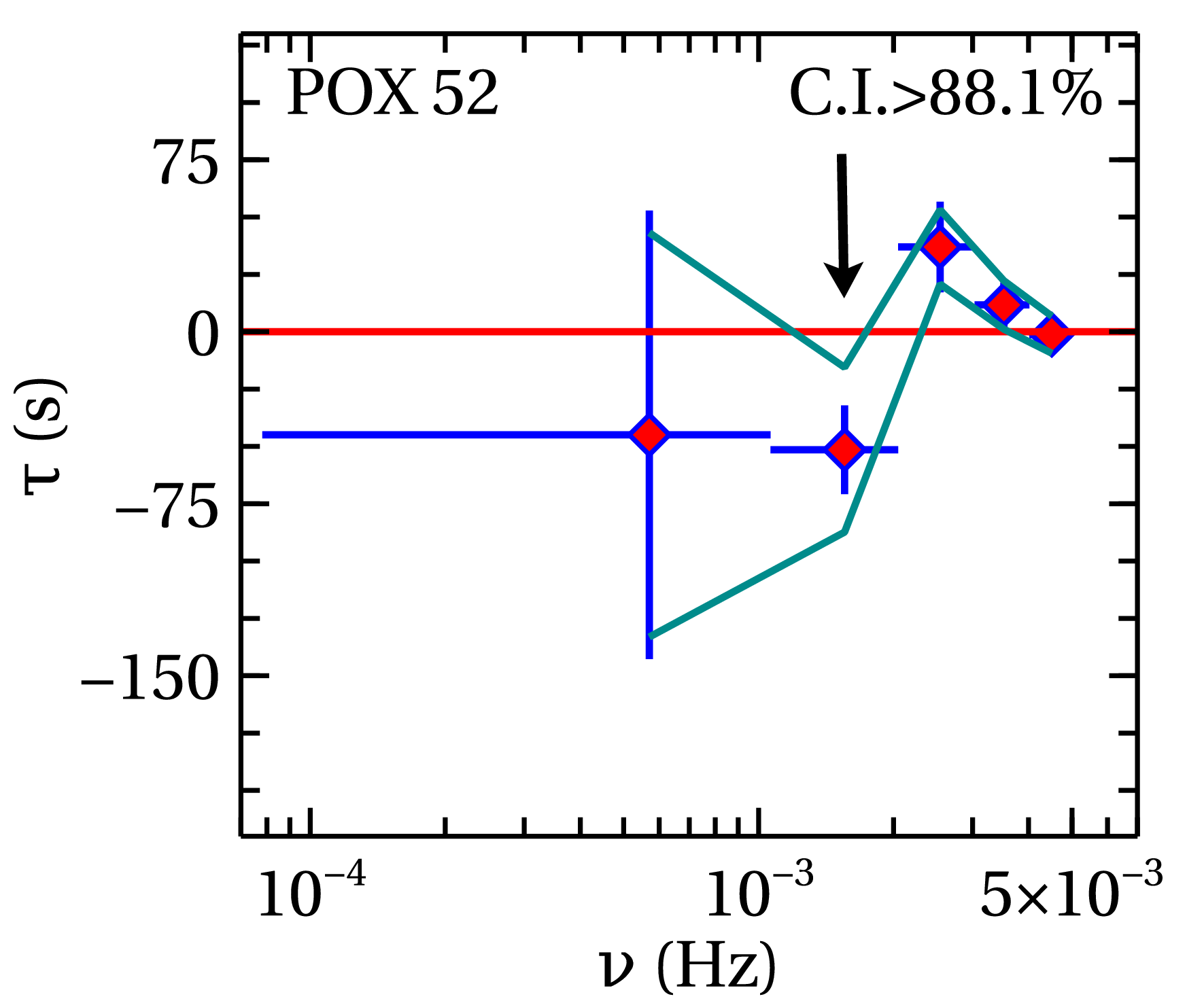}
\caption{Lag-frequency spectra for the sample. The time lag was estimated between the 0.3$-$1\keV{} and 1$-$4\keV{} bands, and negative lag means the soft band lags behind the hard band. The 1$\sigma$ error bars were calculated using the formula given in \citet{no99}. The simulated 1$\sigma$ contour plots are shown by solid green lines. The black arrow refers to the frequency at which the soft lag was observed. C.I. on each panel represents the confidence interval of negative soft lag as determined from Monte Carlo simulations. 6 out of 8 AGNs show soft lags with $>90\%$ confidence.}
\end{center}
\label{fig1}
\end{figure*}

\begin{figure*}
\centering
\begin{center}
\includegraphics[scale=0.4,angle=-0]{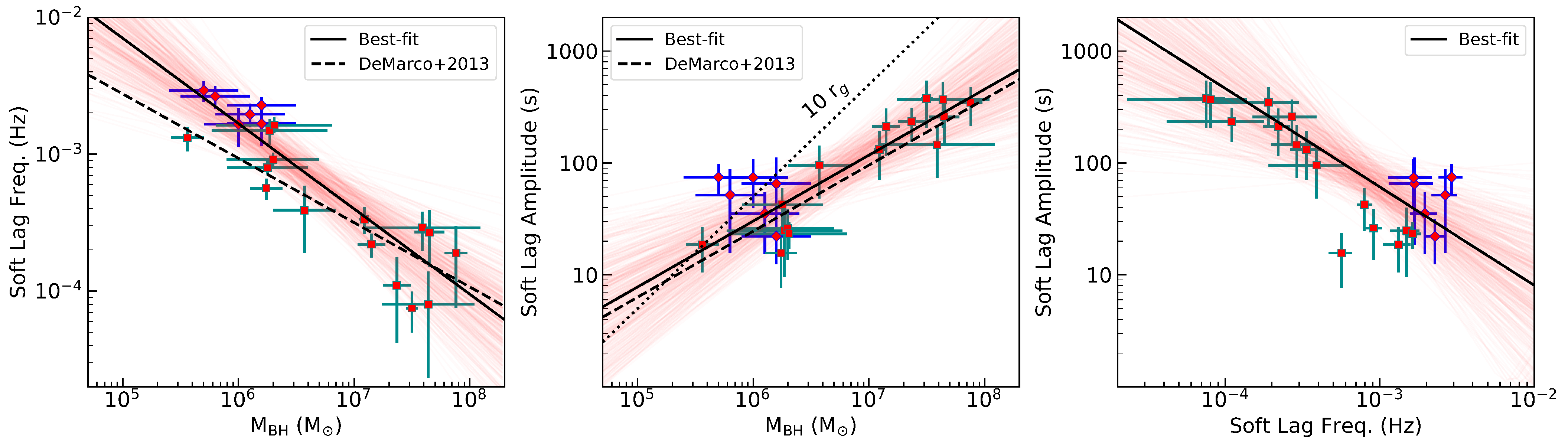}
\caption{Soft lag frequency vs. BH mass (left), soft lag amplitude vs. BH mass (middle), and soft lag amplitude vs. frequency (right). The 1$\sigma$ error bars were determined from Monte Carlo simulations. Sources with a soft lag detected at $>90\%$ confidence are included and marked as diamonds. Sources obtained from DM13 are marked as squares. The dashed lines represent the DM13 best-fit relations. The black solid lines and red shaded regions represent the best-fit linear regression models and corresponding confidence area for the combined sample consisting of the 6 Seyfert~1 galaxies from this work and 15 Seyfert~1 galaxies from DM13. The dotted line represents the light-crossing time at $10 r_{\rm g}$ as a function of mass. The lag amplitudes and frequencies are only corrected for redshift in order to be compared with the DM13 lag data.}
\end{center}
\label{fig2}
\end{figure*}

\section{Frequency-resolved time lag analysis}
\label{sec3}
We measured the time lag as a function of temporal frequency between the soft and hard X-ray light curves following the Fourier method presented in \citet{no99} and \citet{ut14}. The soft and hard X-ray light curves represent the variations of the 0.3$-$1\keV{} and 1$-$4\keV{} energy bands, which are mostly dominated by soft X-ray excess and primary power-law continuum, respectively, and have comparable average count rates. First, we compute the Fourier transforms $S(\nu)$, $H(\nu)$ of two evenly sampled light curves, $s(t)$ and $h(t)$. Then we estimate the cross spectrum by multiplying the Fourier transform of one light curve with the complex conjugate of the Fourier transform of the other light curve. The cross spectrum is defined as $C(\nu)=S^{*}(\nu)H(\nu)=|S(\nu)||H(\nu)|e^{i[\phi_{h}(\nu)-\phi_{s}(\nu)]}$. The time lag between the soft and hard X-ray light curves is derived from the formula, $\tau(\nu)=\phi(\nu)/2\pi \nu$, where $\phi(\nu)={\rm arg}(\langle C(\nu) \rangle)$ is the phase of the average cross spectrum which is estimated by first taking the average of the cross spectral values $C(\nu)$ over multiple, non-overlapping segments and then binned in logarithmically spaced frequency bins. The number of bins is decided by the data quality and exposure time of the observations. The lower frequency bound is set by the inverse of the segment length, and the upper frequency bound is equal to 1/2$\Delta t$, where $\Delta t=100$\s{} is the time bin size used. The errors on the lag were estimated using the formula given in \citet{no99}. In Figure~\textcolor{blue}{A1}, we show the Poisson noise corrected intrinsic variability power of each source as a function of frequency both in the soft and hard X-ray bands. The Poisson noise generally dominated $\geq 4\times10^{-3}$\hz{}, which is well above the frequencies where the soft and hard lags were detected. We verified the significance of lags following the approach of \citet{zo10}. The basic idea of this approach is to generate a pair of artificial light curves, then impose the observed lag between the light curve pair, and check how well the lag can be recovered. We simulated 1000 pairs of soft and hard X-ray Monte Carlo light curves using the method of \citet{em13} after imposing the observed frequency-dependent phase lag between each light curve pair. The resulting simulated light curves have the same statistical and variability properties as the observed pairs. We then derived the lag-frequency spectrum for each pair of the simulated light curves. The confidence interval was estimated from the distribution of the resulting lag values in each frequency bin. The observed lag-frequency spectra and simulated 1$\sigma$ confidence contours for the sample are shown in Figure~\textcolor{blue}{1}. The simulations suggest that the observed 1$\sigma$ error bars were underestimated for a few AGNs due to the data having low counts. The negative soft lags are detected in 6/8 AGNs at the $>90\%$ confidence level, while 5 out of 8 AGNs show positive hard lags with $>94\%$ confidence. From a statistical point of view, the measurement of soft lags in low-mass sources is more challenging as the characteristic time scales are shorter, and thus the number of collected photons is lower. Therefore, the detection significance of soft lags is expected to be lower in the case of lower-mass AGNs. One possible way to raise the detection significance is to obtain longer monitoring of these AGNs. The frequencies, amplitudes, and significance levels of negative soft and positive hard lags are reported in Table~\textcolor{blue}{2}. The soft lag amplitude refers to the largest amplitude negative lag and the corresponding frequency is identified as the soft lag frequency. To check the reality of the soft lag dip, we fit the lag-frequency spectra with a monotonically declining hard lag model, $\tau_{h}=k(\nu/\nu_{0})^{-\alpha}$. The model resulted in poor fits with $\chi^{2}$/d.o.f $>2$ in all sources, which confirms that the hard lag model alone cannot describe the lag profile, and therefore, there is a need for an additional soft lag component (see Fig.~\textcolor{blue}{A2}). The results of the model fitting are shown in Table~\textcolor{blue}{3}. We then checked if a constant hard lag with the observed amplitude can produce phase wrapping at the frequency where we detect the soft lag. The phase-wrapping is a type of aliasing that wraps around from $\pi$ to $-\pi$, and the hard lag can become negative at an interval of $\nu=\frac{n}{2\tau_{\rm h}}$ \citep{ut14}, where $n$ is an odd integer and $\tau_{\rm h}$ is the hard lag amplitude. We verified that the detected soft lag frequencies are well below the phase-wrapping frequencies for six sources in the sample showing soft lags with significance $>90\%$. This confirms that the soft lag was not produced due to the phase wrapping of the low-frequency hard lag in these AGNs.

\section{Results and Discussion}
\label{sec4}
The nature of the lag-frequency spectra is similar to what we usually observe in higher-mass AGNs, as reported by DM13. They show a transition from a positive lag at low frequencies to a negative lag, predicted from reverberation at high frequencies. The lag spectra are typically modeled as a combination of `transfer functions' for the reprocessed reflection component (including all general relativistic effects) and a power-law like frequency response for the intrinsic component (e.g., \citealt{em14,al20}). A detailed transfer function modelling of these data will be shown in a follow up paper. 

At higher frequencies, the soft band variations are delayed relative to the hard band. The soft lag amplitude and frequency correspond to the maximum, negative point in the lag-frequency profile. We detected soft lags with $>90\%$ confidence in 6 out of 8 AGNs. In order to test the dependence of BH mass ($M_{\rm BH}$) on the soft lag frequency ($\nu_{\rm soft}$) and amplitude ($|\tau_{\rm soft}|$), we examined the correlations between $\nu_{\rm soft}$ and $M_{\rm BH}$ as well as the correlations between $|\tau_{\rm soft}|$ and $M_{\rm BH}$ for the combined sample consisting of the 6 Seyfert~1 galaxies from this work and 15 Seyfert~1 galaxies from DM13. We also investigated whether there is a correlation between $|\tau_{\rm soft}|$ and $\nu_{\rm soft}$. We considered the combined sample and applied a Bayesian approach to linear regression with error both in X and Y coordinates \citep{ke07}. The best-fit relations are:
\begin{equation}
{\rm log_{10}}\Big[\frac{\nu_{\rm soft}}{\hz{}}\Big]=-0.62[\pm 0.15] {\rm log_{10}}\Big[\frac{M_{\rm BH}}{10^{7}M_{\odot}}\Big]-3.40[\pm 0.12].
\end{equation}
\begin{equation}
{\rm log_{10}}\Big[\frac{|\tau_{\rm soft}|}{\rm {sec}}\Big]=0.59[\pm 0.19] {\rm log_{10}}\Big[\frac{M_{{\rm BH}}}{10^{7}M_{\odot}}\Big]+2.07[\pm 0.13].
\end{equation}
\begin{equation}
{\rm log_{10}}\Big[\frac{|\tau_{\rm soft}|}{\rm {sec}}\Big]=-0.88[\pm 0.38] {\rm log_{10}}\Big[\frac{\nu_{\rm soft}}{\hz{}}\Big]-0.85[\pm 1.19].
\end{equation}
where $M_{\rm BH}$ is the BH mass in units of $M_{\rm \odot}$, $\nu_{\rm soft}$ and $|\tau_{\rm soft}|$ are the soft lag frequency and amplitude in Hz and second, respectively. The soft lag frequency vs. BH mass, soft lag amplitude vs.\ BH mass, and soft lag amplitude vs. frequency plots and their corresponding best-fit linear models in log-log space are shown in Figure~\textcolor{blue}{2}. The results are consistent with the mass-lag scaling relations of DM13. The spread in the lag amplitude may be the result of dilution effects. For some AGNs, the scatter could be due to the changes in the coronal geometry (e.g., \citealt{al20}).

We also performed a Spearman's rank correlation test on the combined sample to evaluate the correlations between these three parameters. The Spearman coefficient values for the $\nu_{\rm soft}-M_{\rm BH}$, $|\tau_{\rm soft}|-M_{\rm BH}$ and $|\tau_{\rm soft}|-\nu_{\rm soft}$ correlations are $-0.87$, $0.73$, and $-0.71$ with null-hypothesis probabilities of $2.9\times 10^{-7}$, $1.7\times 10^{-4}$, and $3.4\times 10^{-4}$, respectively, suggesting that the inverse correlation between lag frequency and BH mass is more significant and secured.

If we assume a lamppost model with a stationary corona above the disc and the coronal height is significantly larger than the innermost stable circular orbit (ISCO), then the soft lag amplitude in flat, Euclidean spacetime can be approximated as the light-crossing time delay between the corona and the face-on disc,
\begin{equation}
\tau_{\rm soft} \approx 50\left(\frac{h_{\rm f}}{r_{\rm g}}\right) \left(\frac{M_{\rm BH}}{10^{7}M_{\rm \odot}}\right),
\end{equation}
where $h_{\rm f}$ is the height of the corona in flat, Euclidean spacetime in units of $r_{\rm g}=GM_{\rm BH}/c^{2}$, $M_{\rm BH}$ is the BH mass in units of $M_{\rm \odot}$. The light-crossing time over $1r_{\rm g}$ for a BH of mass $10^{7}M_{\rm \odot}$ is 50\s{}. However, the measured lags get diluted as the reflection-dominated soft band (0.3$-$1\keV{}) contains a fraction of photons from the primary continuum and the hard band (1$-$4\keV{}) dominated by the direct continuum includes a fraction of photons from the reflected emission (e.g., \citealt{wf13}). Therefore, to make an accurate measurement of the coronal geometry, we determined the dilution fractions from the modeling of source photon count spectra. All these sources have a soft X-ray excess below around 1\keV{}, which can be described by a relativistic reflection model. For the 6 Seyfert~1 AGNs where the soft lag detection significance exceeded $90\%$, we fit the 0.3$-$10\keV{} spectral data with the model, [\textsc{tbabs$\times$(relxill+zpowerlw)}], and estimated the direct and reflected photon flux both in the soft and hard X-ray bands. We then calculated the direct-to-reflected flux fraction ($\delta_{\rm soft}$) in the soft band and the reflected-to-direct flux fraction ($\delta_{\rm hard}$) in the hard band (see Table~\textcolor{blue}{2}). The observed soft and hard lags get diluted by a factor of $\frac{\delta_{\rm soft}}{1+\delta_{\rm soft}}$ and $\frac{\delta_{\rm hard}}{1+\delta_{\rm hard}}$, respectively \citep{ut14}. Therefore, to obtain the dilution-corrected lag times, we divided the measured soft and hard lags by the dilution factors, $\frac{\delta_{\rm soft}}{1+\delta_{\rm soft}}$ and $\frac{\delta_{\rm hard}}{1+\delta_{\rm hard}}$, respectively.

In reality, however, when light propagates close to the black hole, it will be delayed (the `Shapiro delay') and hence photon takes longer than 100\s{} to cross $2r_{\rm g}$ close to a black hole of mass $10^{7}M_{\rm \odot}$. Therefore, it is necessary to not only take in the dilution effects which shorten the lags from their intrinsic values, but also the Shapiro delays close to the black hole which lengthen the lags based on the simple estimate of equation~\textcolor{blue}{(4)}. Figure~\textcolor{blue}{3} shows soft reverberation time lags calculated assuming a stationary corona at different heights on the rotation axis of the black hole in curved spacetime. The time lag in the curved spacetime is calculated using the X-ray reverberation model described in \citet{wi16} and corresponds to the average time lag integrated across the whole disc from a point source at different heights. The model accounts for all of the light paths from the corona to the disc to an observer and the delay of each ray as it passes close to the black hole. The height of the corona, $h_{\rm c}$, as shown in Table~\textcolor{blue}{2}, is corrected for Shapiro delays, dilution effects, and source redshift. We find that the corona of the sample extends at an average height of around $10 r_{\rm g}$ on the BH rotation axis, which is in agreement with the characteristic scale height of the higher-mass AGN coronae \citep{em14,wi19}. It is worth mentioning that timing results have previously preferred higher source heights than inferred from spectroscopy \citep{al20}, which could be due to the lamppost assumption breaking down in a scenario where the average reflection spectrum is dominated by a constant low-height part of the corona, while the variability is dominated by a flickering high-height part. It is also possible that comptonization of the reflection spectrum acts to further smooth the soft excess produced by reflection \citep{wg15,ba20}. This could lead us to measure more extreme blurring from spectroscopy than from the timing.

\begin{figure}
\centering
\begin{center}
\includegraphics[scale=0.52,angle=-0]{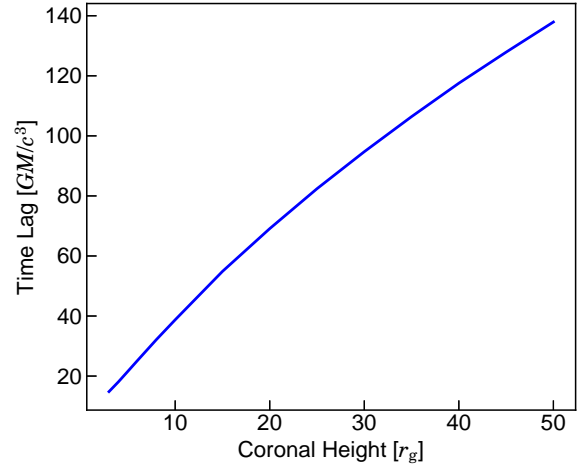}
\caption{Soft reverberation time lags due to a stationary, isotropic corona on the black hole rotation axis at different heights in curved spacetime. We calculated time lags using the reverberation model of \citet{wi16}, which accounts for all of the light paths from the corona to the disc to the observer and the `Shapiro delay'.}
\end{center}
\label{fig2b}
\end{figure}

At the lowest observed frequencies of $\sim[3-8]\times 10^{-4}$\hz{}, the hard band lags behind the soft band variations. We explain the hard lag in the context of viscous propagation of mass accretion rate fluctuations propagating inwards through the accretion disc to create variations in the corona \citep{ko01,hogg16}. Therefore, the hard lag ($\tau_{\rm hard}$) can correspond to the viscous timescale ($\tau_{\rm vis}$) of the variations that originate at the edge of the corona and propagate inwards, 
\begin{equation}
\tau_{\rm hard} = \tau_{\rm vis} = 500\left(\frac{r_{\rm c}}{h_{\rm c}}\right)^{2} \left(\frac{M_{\rm BH}}{10^{7}M_{\rm \odot}}\right)\left(\frac{r_{\rm c}}{r_{\rm g}}\right)^{3/2},
\end{equation}
where the viscosity parameter, $\alpha\approx 0.1$. $M_{\rm BH}$ is the BH mass in units of $M_{\rm \odot}$. $r_{\rm c}$ and $h_{\rm c}$ represent the radius and height of the corona in units of $r_{\rm g}$, respectively. From the measured hard lag and source height, we estimated the radial extent of the corona in the context of propagating fluctuation models using equation~\textcolor{blue}{(5)} after correcting for dilution effects and source redshift. The radial extent of the corona for each source in the sample is listed in Table~\textcolor{blue}{2}. We find that the corona extends at an average radius of $r_{\rm c} \sim 6r_{\rm g}$ over the surface of the accretion disc on an average timescale of $\sim30$ minutes. The radial extent of the corona is in agreement with the break radius measured from the disk emissivity profile modeling \citep{wg15} of higher-mass AGNs. This result is also consistent with X-ray microlensing observations \citep{ch09,ch12}. We note that the measured hard lags are not well sampled in these data sets because of the limited frequency window. They might extend to much lower frequencies and larger amplitudes (see \citealt{pa19}).

\section{Conclusions}
\label{sec5}
We report the first results on the measurement and analysis of time lags between the soft (0.3$-$1\keV{}) and hard (1$-$4\keV{}) X-ray bands in the Fourier domain for a sample of low-mass ($M_{\rm BH}<3\times 10^{6} M_{\rm \odot}$) active galaxies. We summarize our conclusions as follows:

\begin{enumerate}
\item The shape of the lag-frequency spectra resembles the lag-frequency profile of higher-mass AGNs with a transition from a hard lag at low frequencies to a soft lag at high frequencies.

\medskip

\item The origin of soft lags can be explained in the context of disc reflection, which reverberates in response to the direct coronal emission. The average coronal height calculated from soft reverberation lags is around $10 r_{\rm g}$, which indicates that the corona of lower-mass AGNs is as compact as their higher-mass counterparts.

\medskip

\item The origin of hard lags is most likely the viscous propagation of fluctuations that originate at the edge of the corona and propagate inwards through it. On an average timescale of around 30 minutes, the inferred average radial extent of the corona on the disc surface is around $6r_{\rm g}$. This value agrees well with the break radius obtained from the disc emissivity profile fitting of higher-mass AGNs.

\medskip

\item The inverse correlation between the soft lag amplitude and frequency implies that shorter timescale variations have been generated in a more compact X-ray emitting region that radiates at the smaller disc radii and hence closer to the central BH. 

\medskip

\item The correlations between the BH mass and the soft-lag amplitude/frequency for a broad mass range ($\log M_{\rm BH}\sim 5-8$) follow the DM13 scaling relations suggesting the presence of a BH mass-independent universal accretion process in accreting black hole systems.

\medskip

\item This work will pave the way for more comprehensive investigations into the low-mass end of AGNs, which is crucial to understand the nature of cosmological `seed' black holes. The overall conclusion is that the existence of time lags between various X-ray energy bands is inevitable unless some strange accretion physics is at work in this low-mass range. We aim to further explore this kind of accreting black hole systems in great detail with future long-duration, multi-wavelength observations.

\end{enumerate}

\section{acknowledgments}
We thank the anonymous reviewer for their time and constructive feedback that improved this work. LM acknowledges support from the Department of Science of Technology (DST), India, fellowship agreement No. PDF/2020/3226. BDM acknowledges support from the European Union's Horizon 2020 research and innovation programme under the Marie Sk{\l}odowska-Curie grant agreement No.\ 798726 and via Ram\'on y Cajal Fellowship RYC2018-025950-I. AGM acknowledges partial support from Narodowym Centrum Nauki (NCN) grants 2016/23/B/ST9/03123 and 2018/31/G/ST9/03224, and also from NASA via NASA-ADAP Award NNX15AE64G. We dedicate this paper to healthcare workers fighting the COVID-19 global pandemic.

This research has made use of the NASA/IPAC Extragalactic Database (NED), which is operated by the Jet Propulsion Laboratory, California Institute of Technology, under contract with the NASA. 

This research has made use of data, software and/or web tools obtained from the High Energy Astrophysics Science Archive Research Center (HEASARC), a service of the Astrophysics Science Division at NASA/GSFC and of the Smithsonian Astrophysical Observatory's High Energy Astrophysics Division.

This research has made use of ISIS functions (ISISscripts) provided by ECAP/Remeis observatory and MIT (\url{http://www.sternwarte.uni-erlangen.de/isis/}).

\section{Data availability}
All data used in this work are publicly available from the \xmm{} science archive (\url{http://nxsa.esac.esa.int/}).


\appendix
\section{Additional Table and plots}
\label{sec:appendix}

\begin{table*}
\caption{Observation log of the used sample. The columns are (1) name of the source, (2) right ascension, (3) declination, (4) \xmm{} observation Id, (5) observation start date, (6) exposure time in ks after excluding the intervals of background flares.}
\begin{center}
\scalebox{0.95}{%
\begin{tabular}{ccccccc}
\hline 
Source Name [Short]   & RA  & DEC     & Obs. Id &   Date        & Filtered          \\
                    & (Deg) & (Deg)   &         & (yyyy-mm-dd)  & duration (ks)  \\  
(1)    & (2)  & (3) & (4) & (5) & (6)  \\                                                      
\hline 
SDSS~J010712.03$+$140844.9 [J0107] & 16.800 &  14.146 & 0305920101 & 2005-07-22  & 34.8   \\ [0.2cm]

SDSS~J094240.92$+$480017.3 [J0942] & 145.6705 & 48.005  & 0201470101 & 2004-10-14  & 42.0         \\ [0.2cm]
                                 &          &         & 0201470301 & 2004-11-13  & 13.0            \\ [0.2cm]
       
SDSS~J102348.44$+$040553.7 [J1023] & 155.952 & 4.098  & 0108670101 & 2000-12-05  & 51.1   \\ [0.2cm]
                                  &          &       & 0605540201 & 2009-12-13  & 111.6   \\ [0.2cm]
                                  &          &       & 0605540301 & 2009-05-08  & 56.7    \\ [0.2cm]

SDSS~J114008.71$+$030711.4 [J1140] & 175.036 & 3.120 & 0305920201 & 2005-12-03  & 39.6  \\ [0.2cm]
                                 &          &         &0724840101 & 2013-12-18  & 38.1   \\ [0.2cm]
                                 &          &         &0724840301 & 2014-01-01  & 73.0   \\ [0.2cm]

SDSS~J134738.23$+$474301.9 [J1347] & 206.909 & 47.717  & 0744220701 & 2014-11-22  & 30.8   \\ [0.2cm]

SDSS~J143450.62$+$033842.5 [J1434] & 218.711 & 3.645  & 0305920401 & 2005-08-18  & 36.3   \\ [0.2cm]
                                 &          &         & 0674810501 & 2011-08-16  & 11.6   \\ [0.2cm]

SDSS~J155909.62$+$350147.4 [J1559] & 239.790 & 35.030   & 0112600801 & 2003-01-16  & 14.6  \\ [0.2cm]
                                 &          &         & 0744290101 & 2015-03-02  & 98.7   \\ [0.2cm]
                                 &          &         & 0744290201 & 2015-02-24  & 95.3  \\ [0.2cm]    

POX52 & 180.737 & $-$20.934 & 0302420101 & 2005-07-08  & 85.3   \\ [0.2cm]
\hline 
\end{tabular}}
\end{center} 
\label{tabA1}           
\end{table*}

\begin{figure*}
\centering
\begin{center}
\includegraphics[scale=0.32,angle=-0]{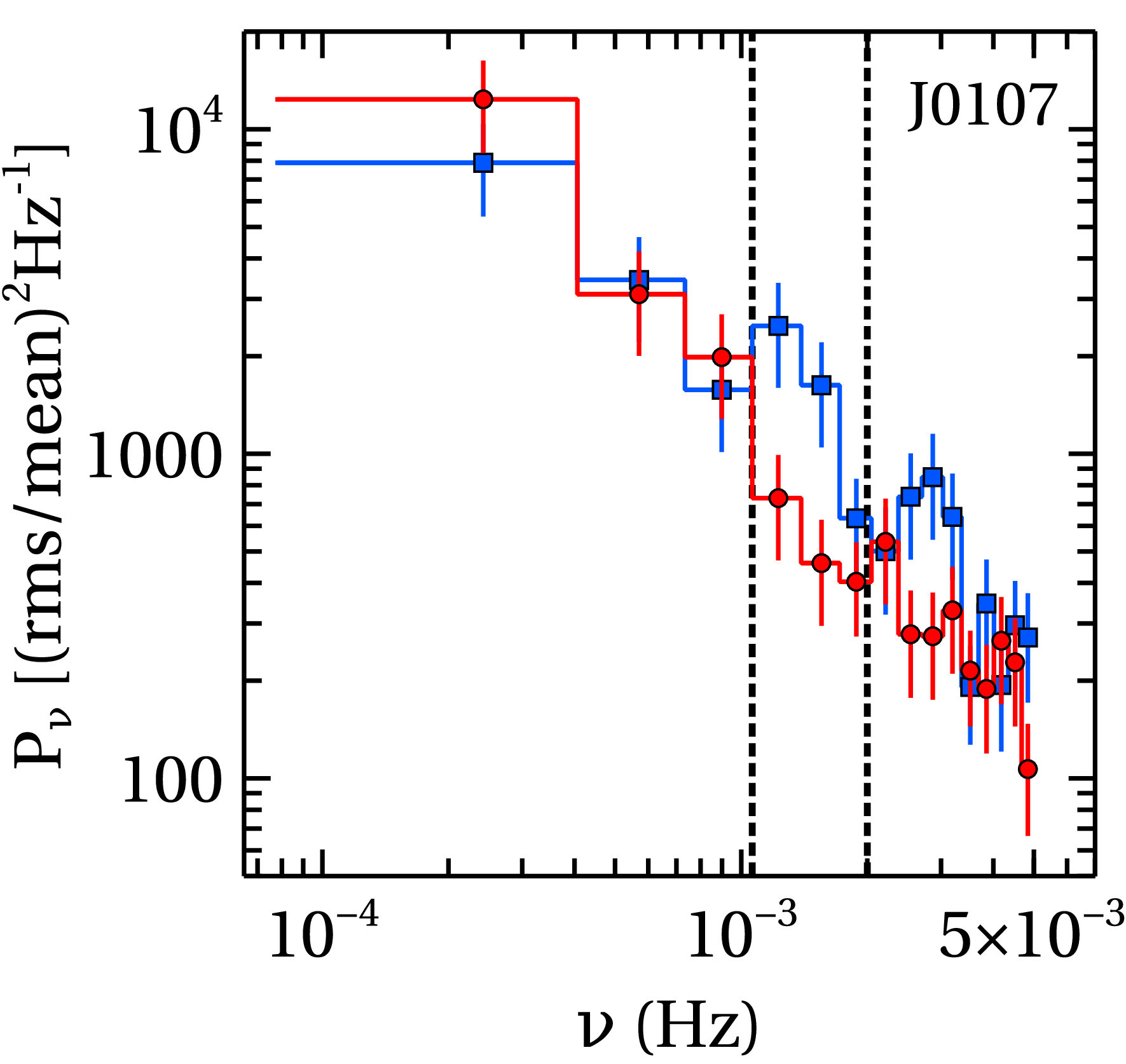}
\includegraphics[scale=0.32,angle=-0]{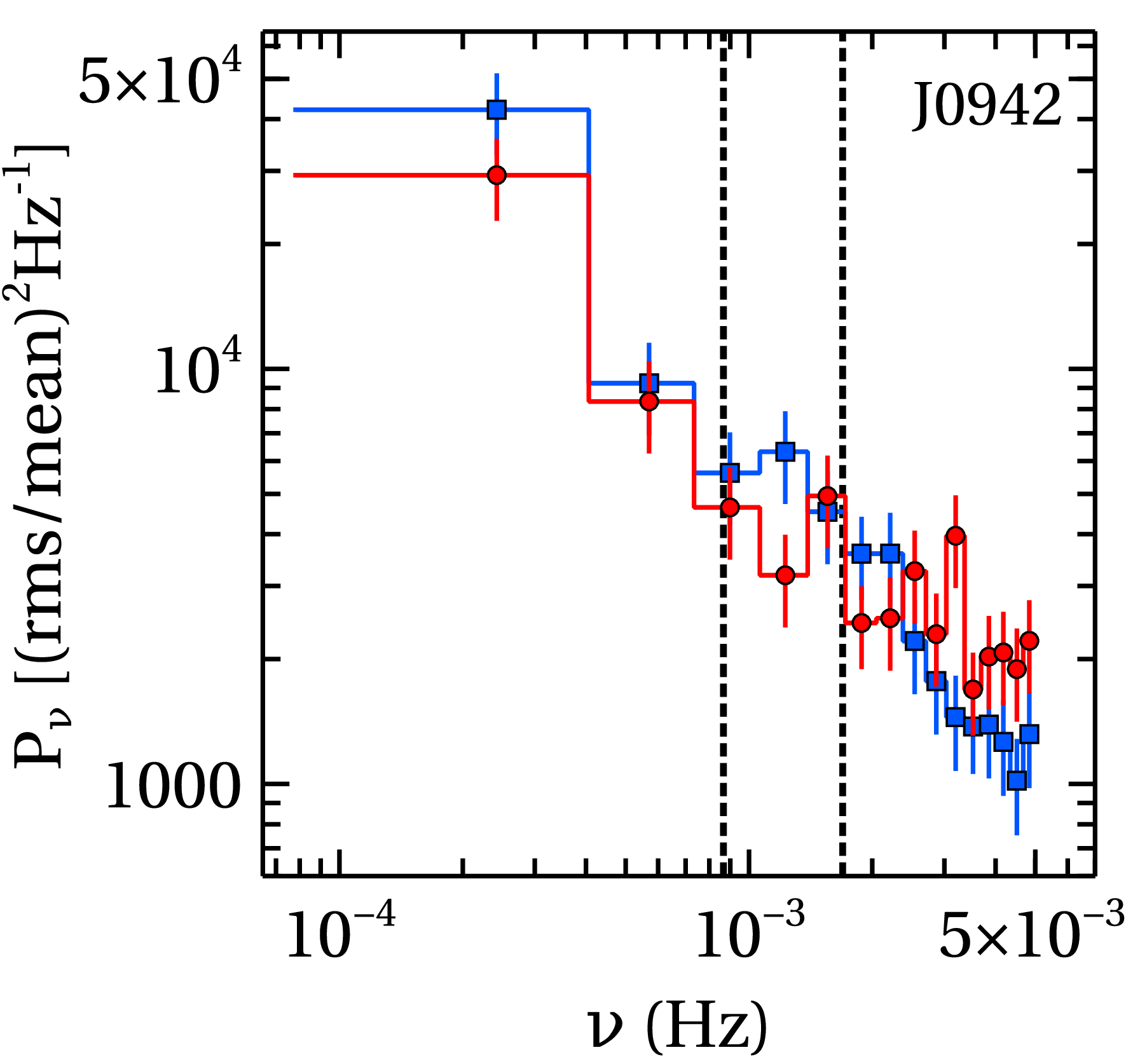}
\includegraphics[scale=0.32,angle=-0]{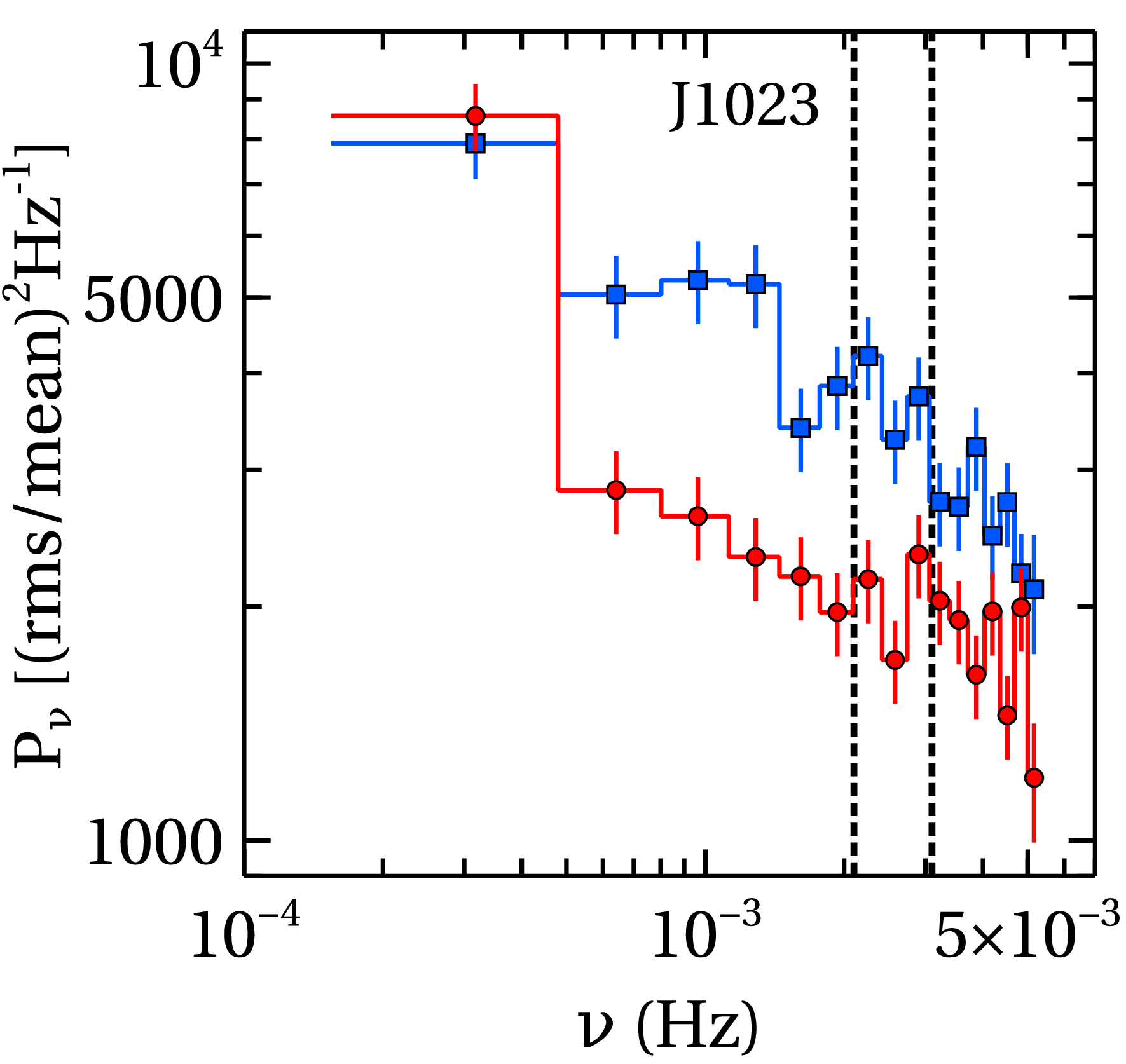}
\includegraphics[scale=0.32,angle=-0]{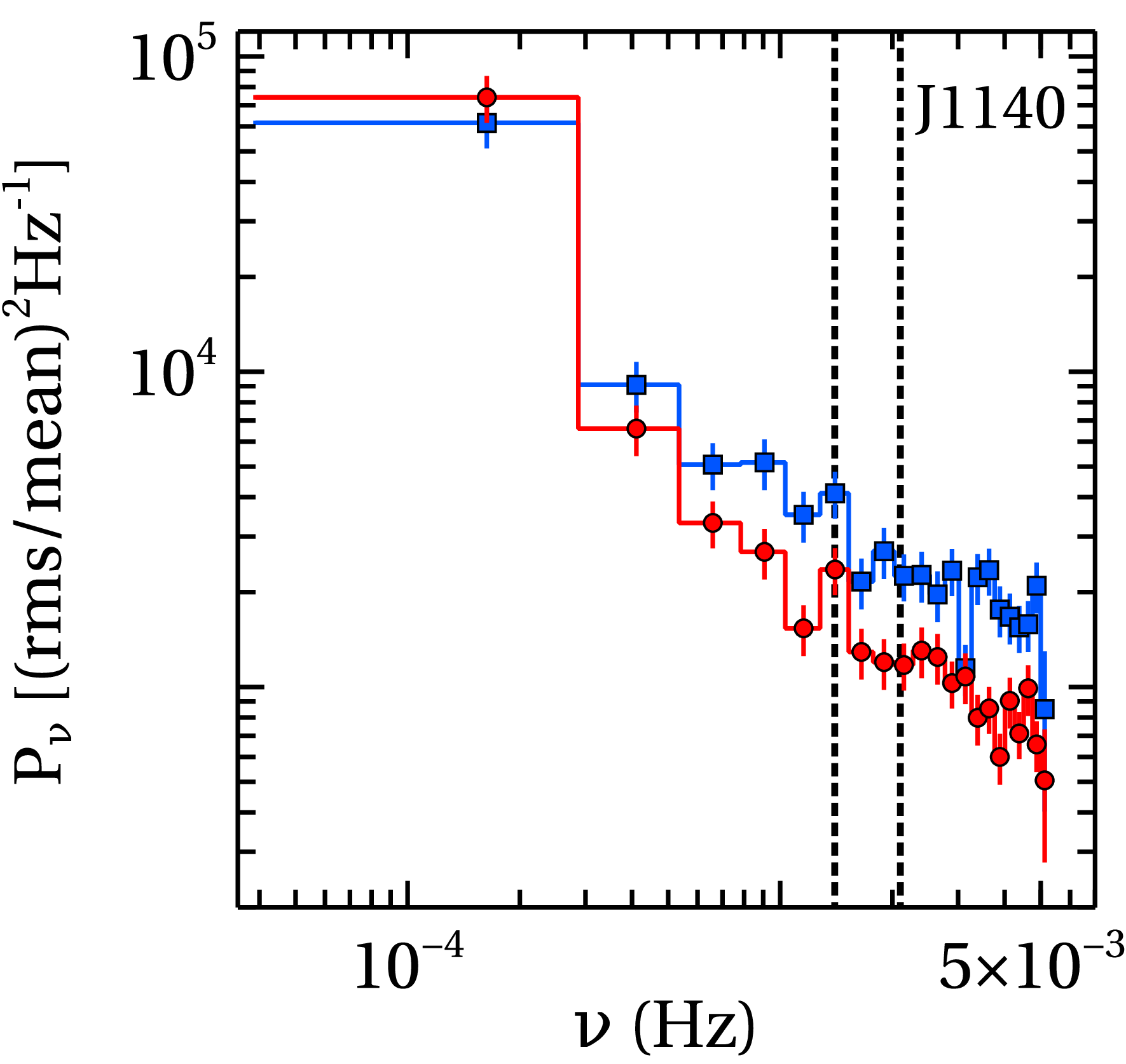}
\includegraphics[scale=0.32,angle=-0]{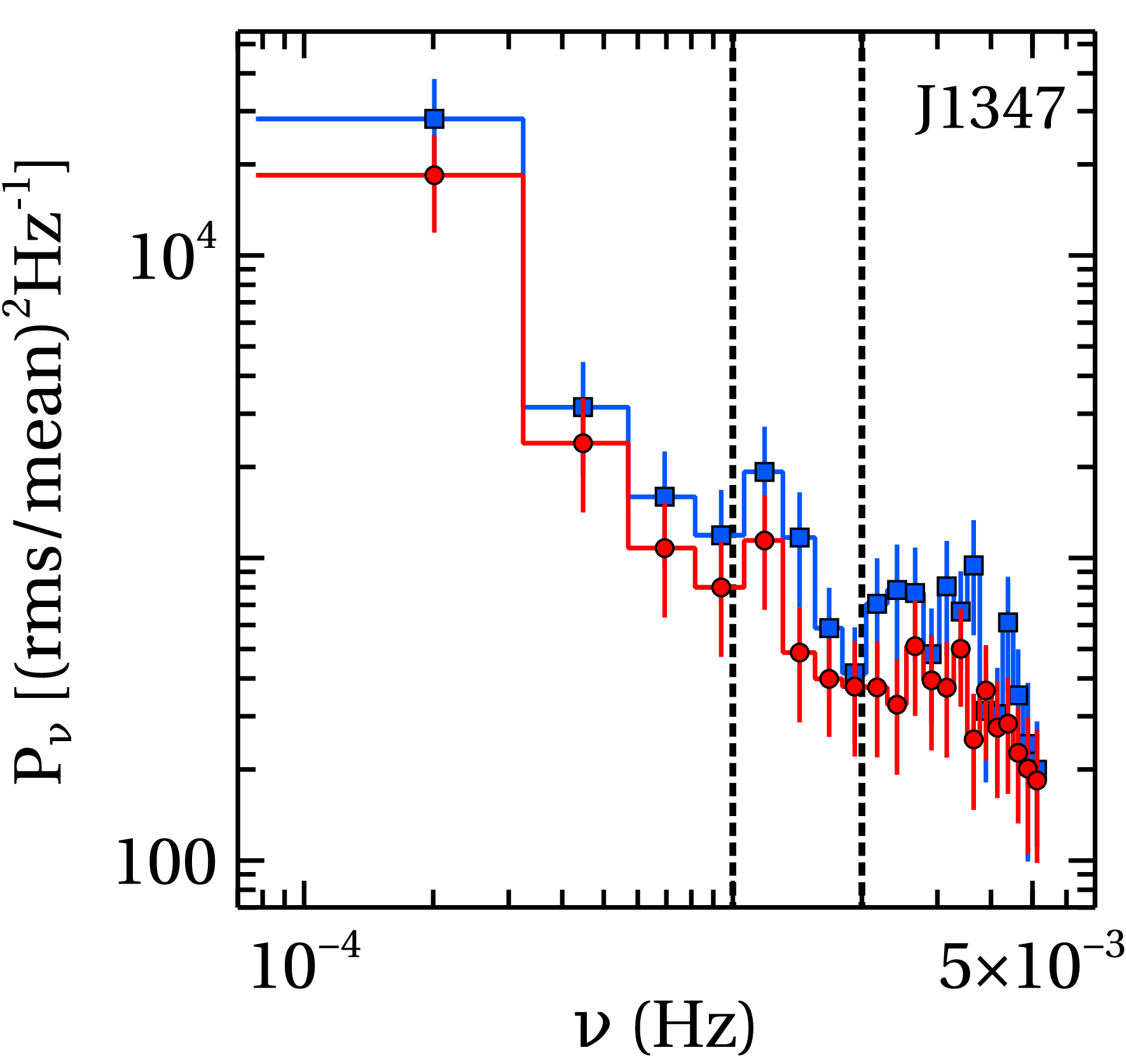}
\includegraphics[scale=0.32,angle=-0]{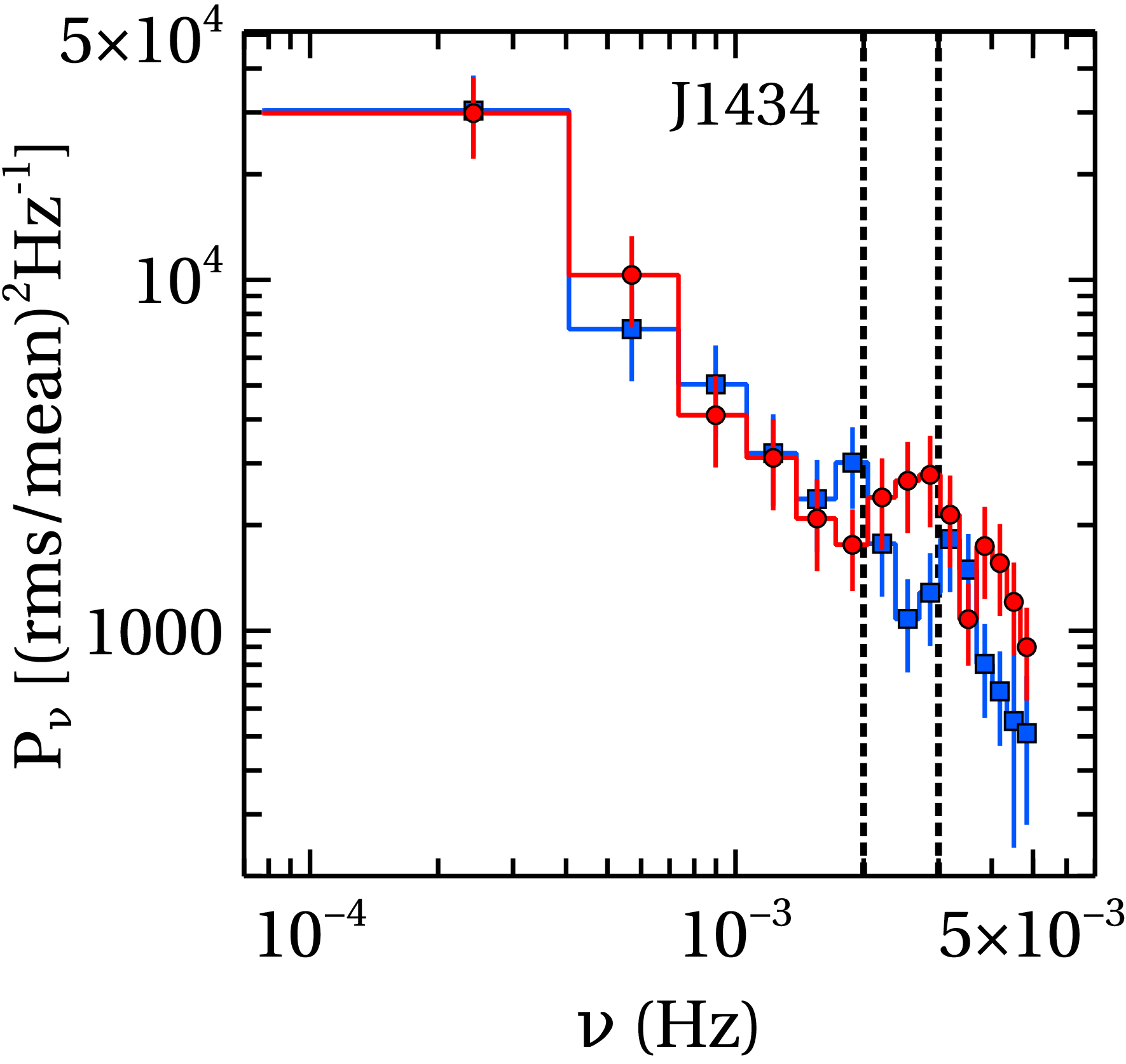}
\includegraphics[scale=0.32,angle=-0]{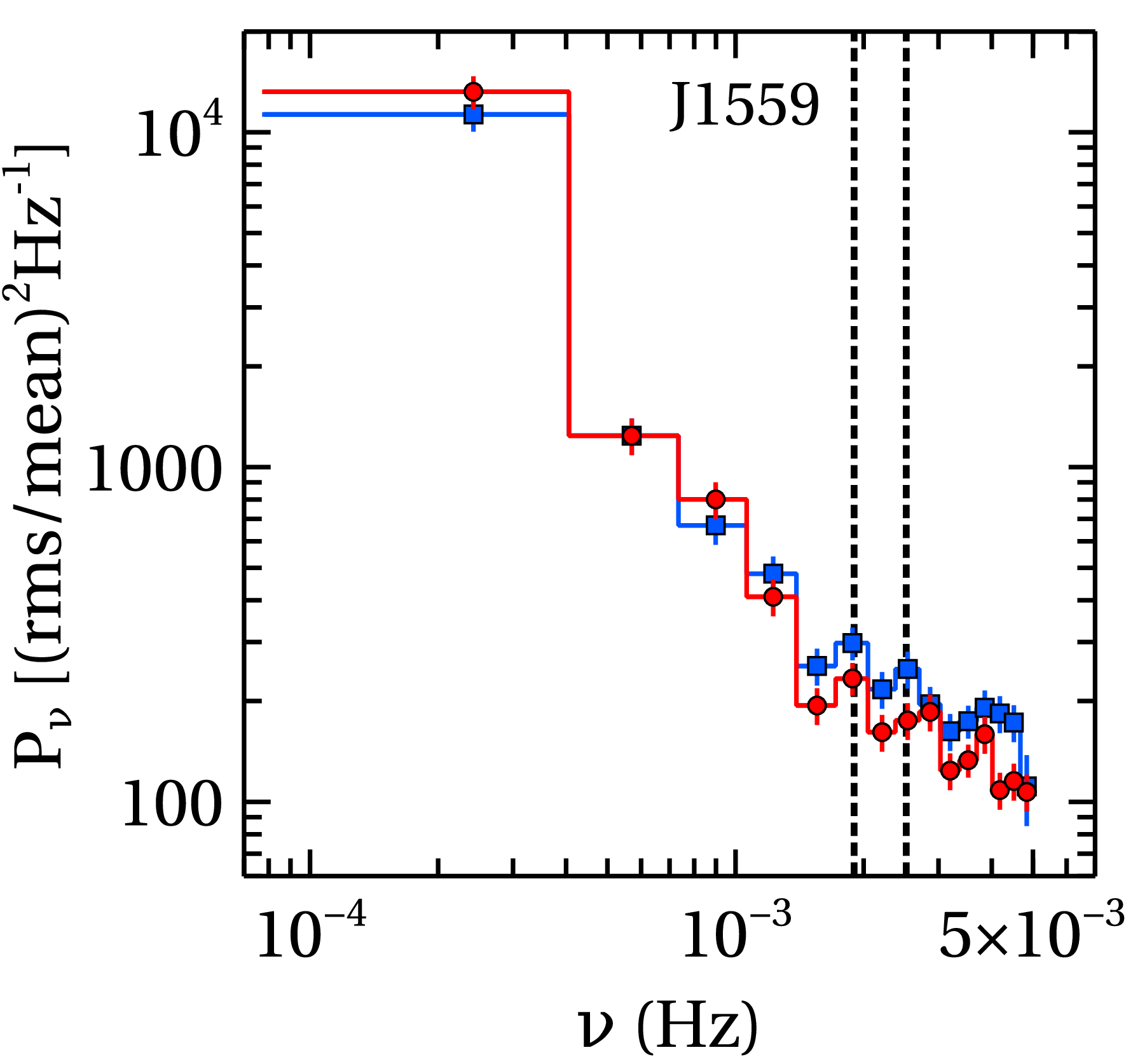}
\includegraphics[scale=0.32,angle=-0]{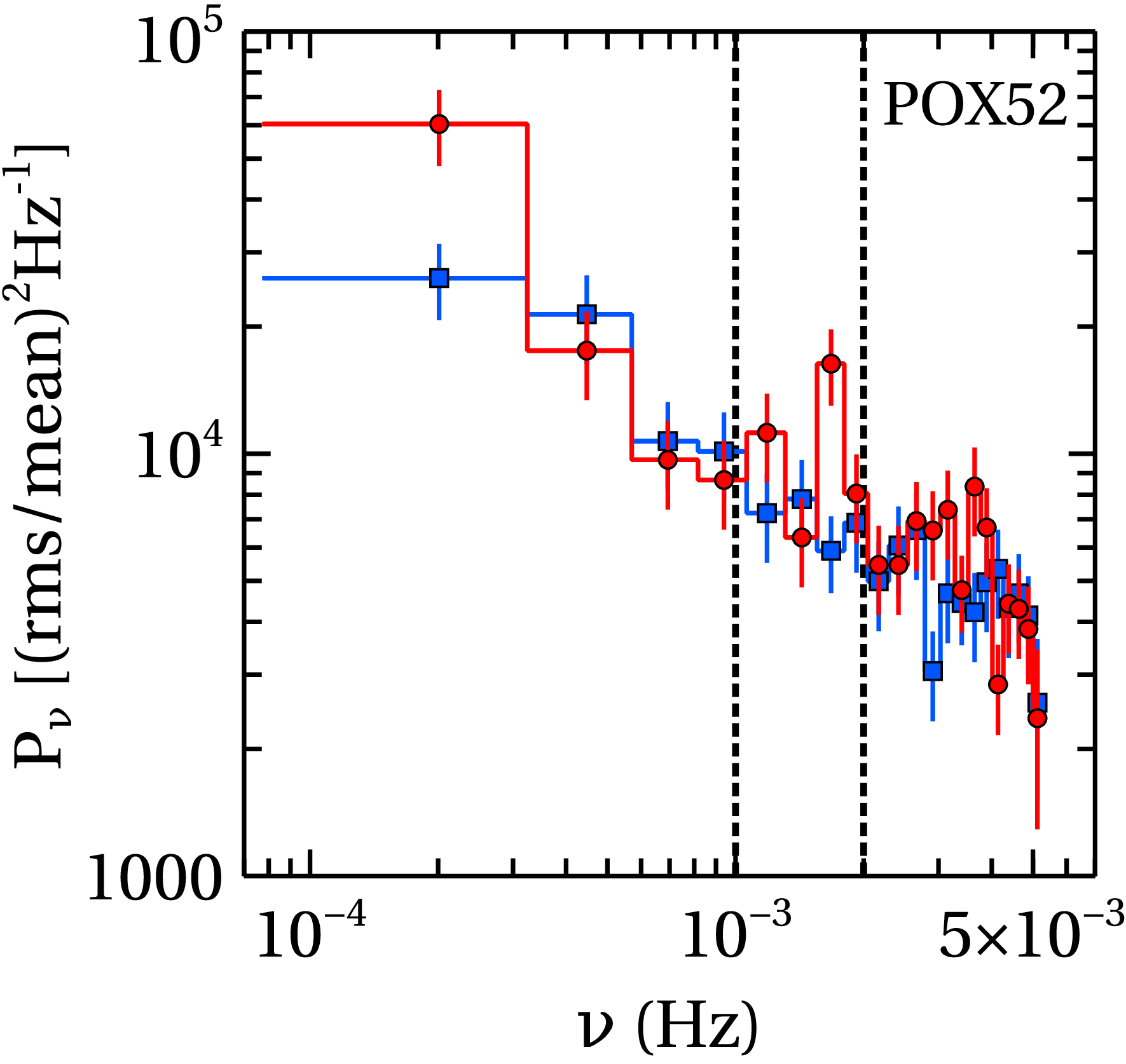}
\caption{Poisson-noise subtracted power spectral density in the soft (red circles) and hard (blue squares) X-ray bands, demonstrating variability power intrinsic to the source as a function of frequency. The vertical dashed lines indicate the range of frequencies where soft lags were detected.}
\end{center}
\label{fig_extra}
\end{figure*}

\begin{figure*}
\centering
\begin{center}
\includegraphics[scale=0.32,angle=-0]{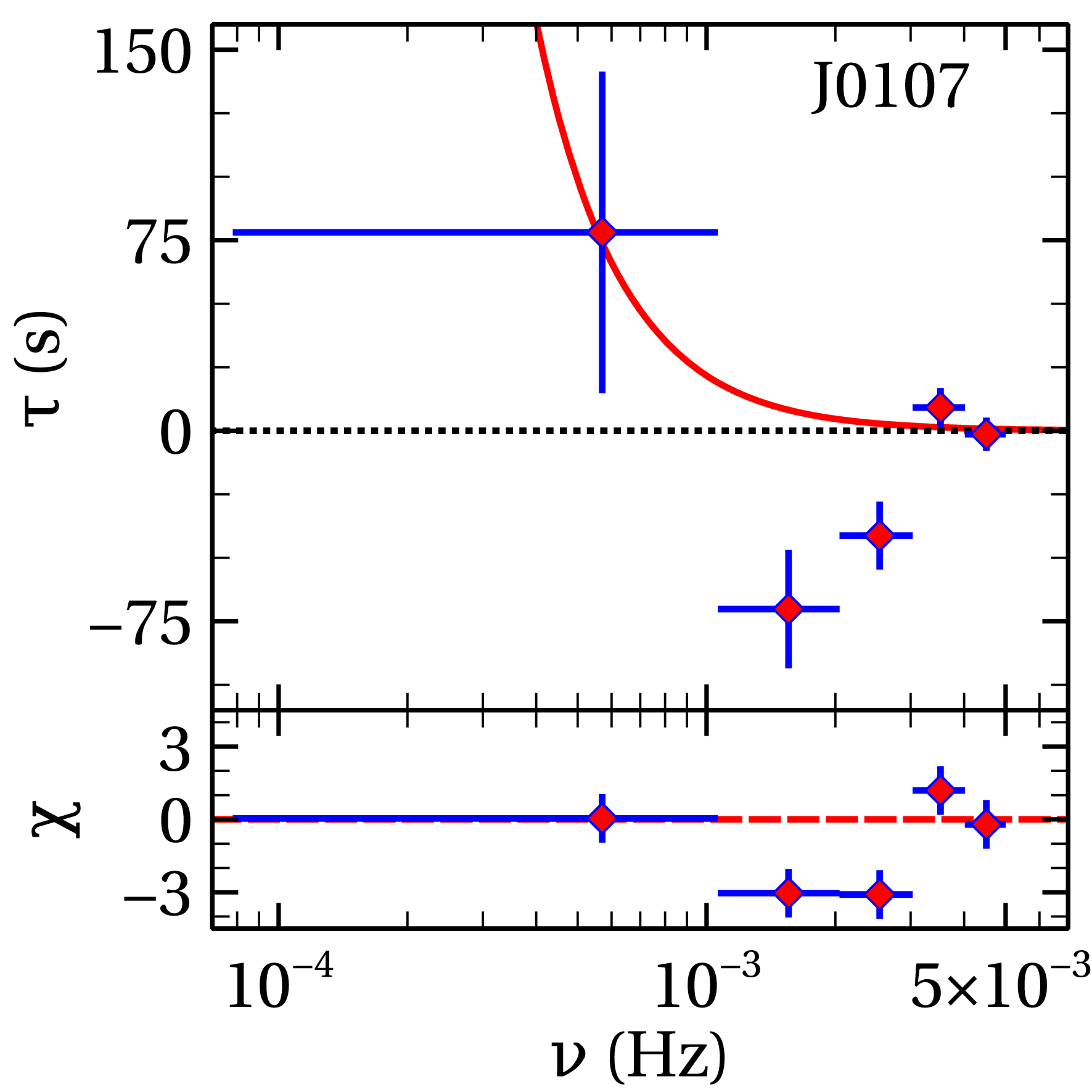}
\includegraphics[scale=0.32,angle=-0]{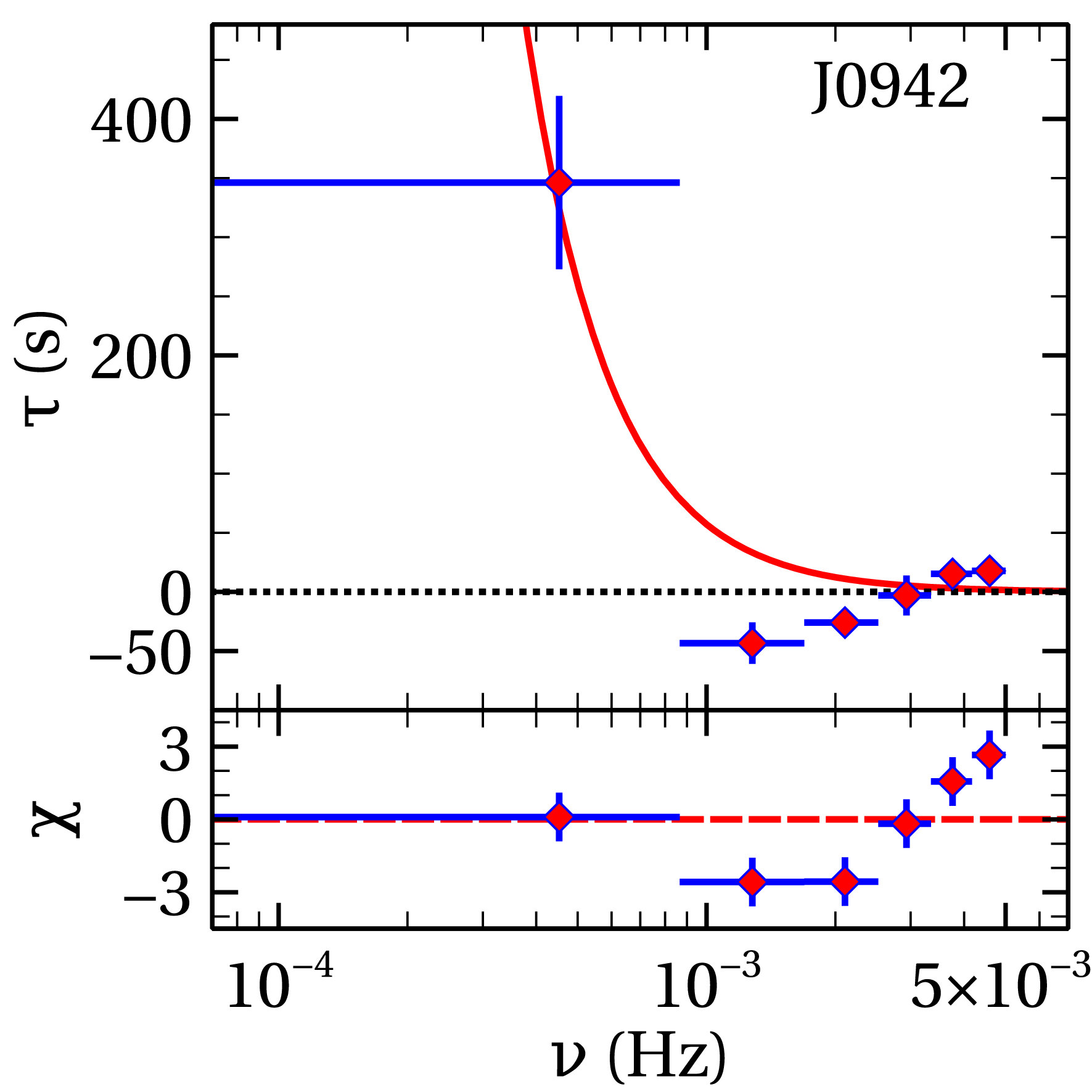}
\includegraphics[scale=0.32,angle=-0]{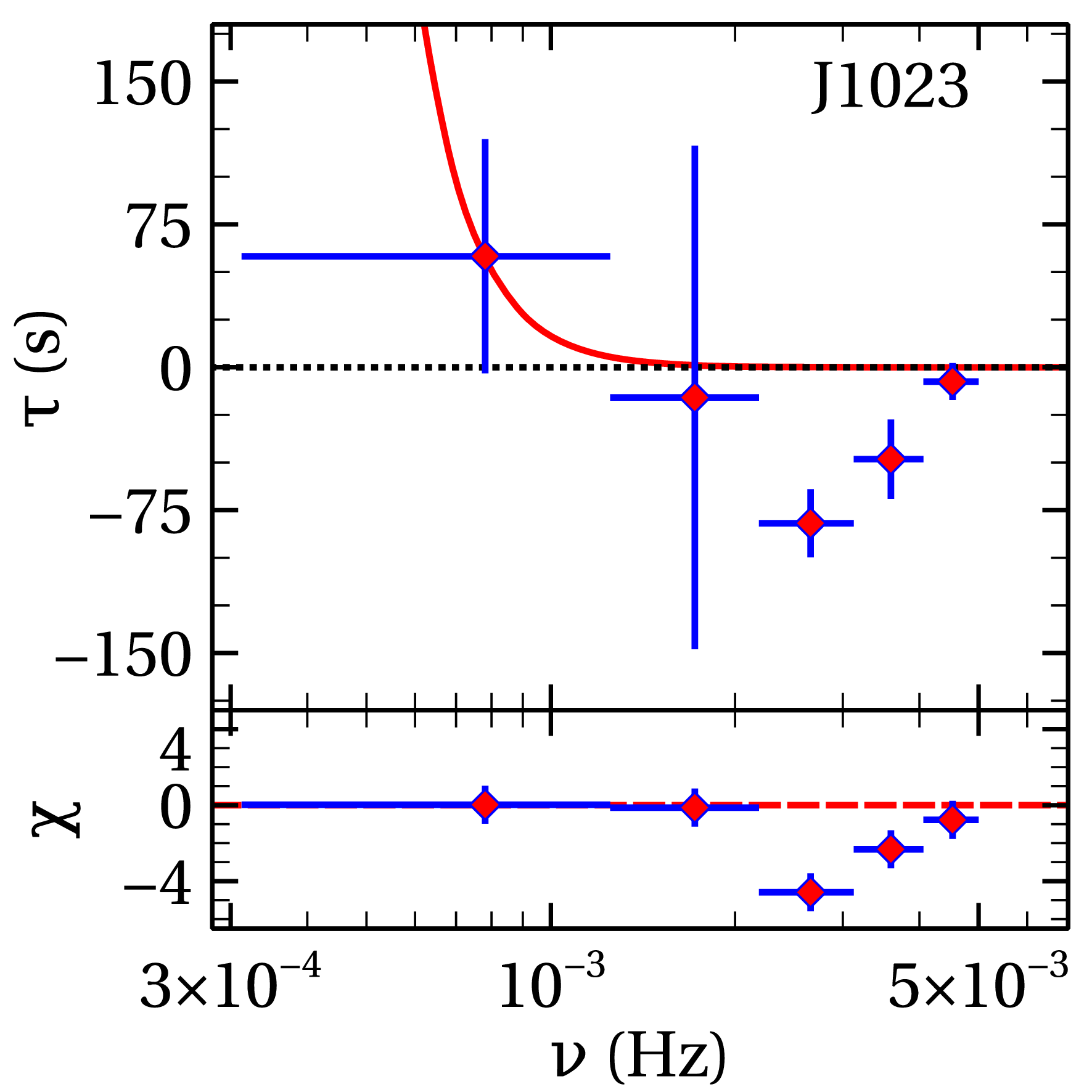}
\includegraphics[scale=0.32,angle=-0]{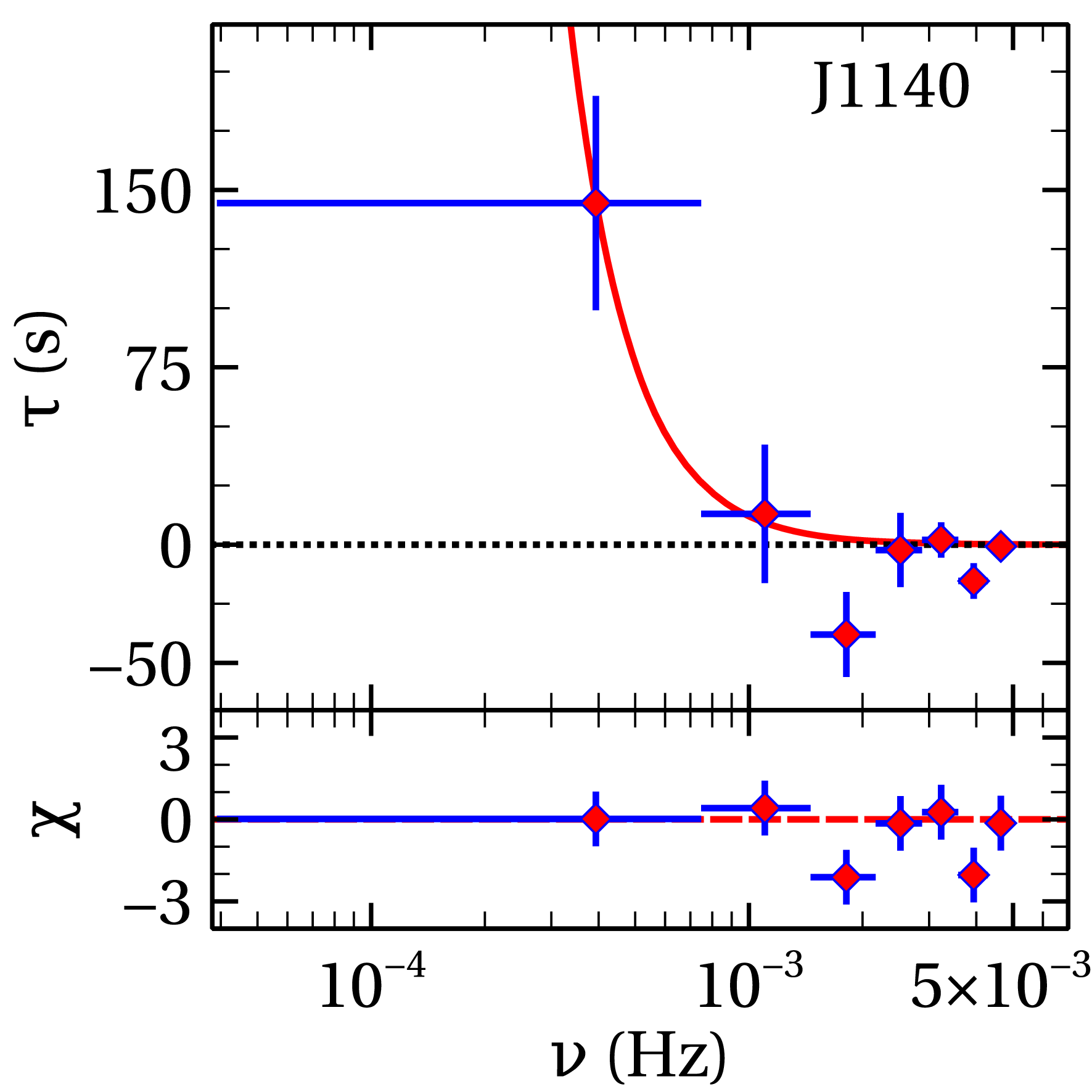}
\includegraphics[scale=0.32,angle=-0]{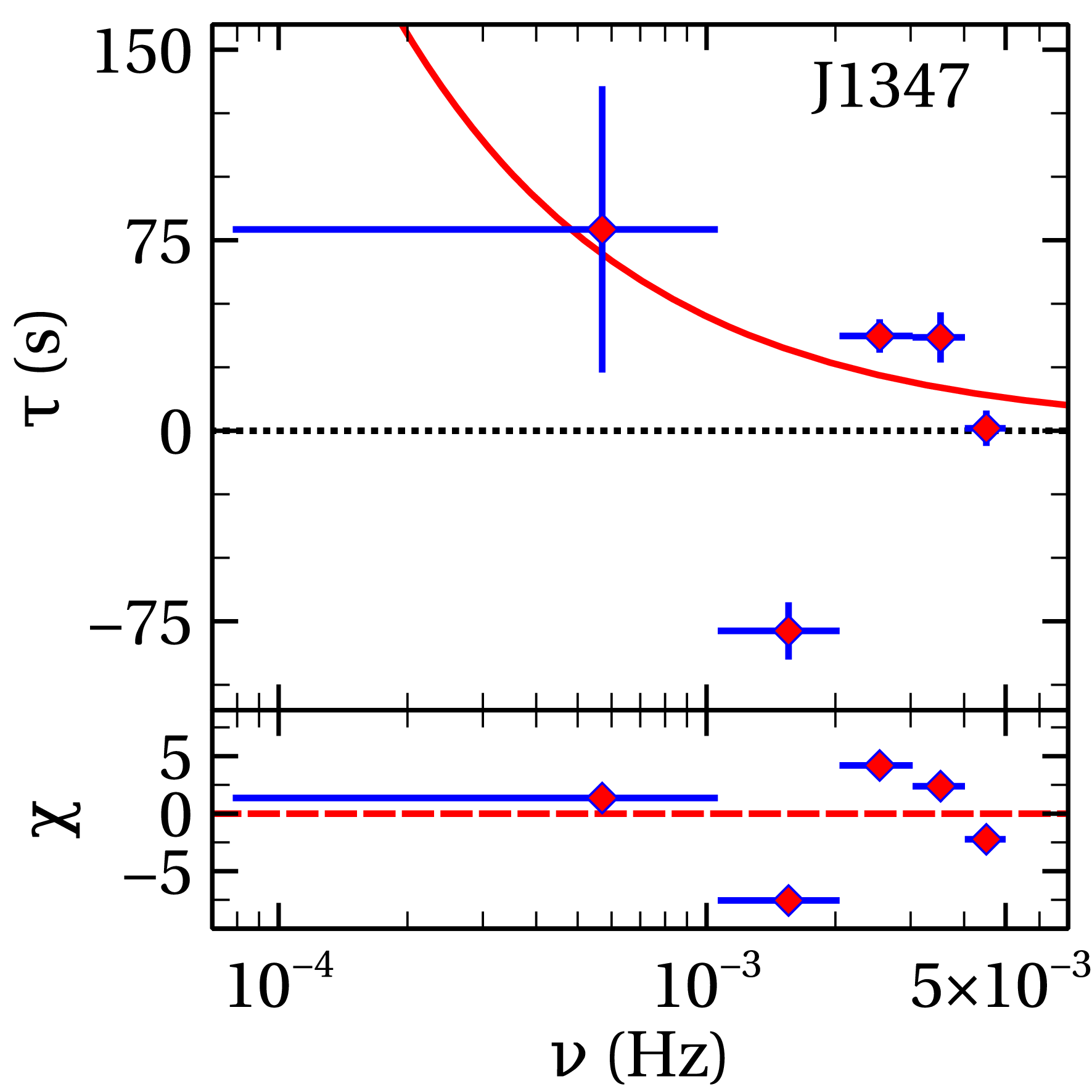}
\includegraphics[scale=0.32,angle=-0]{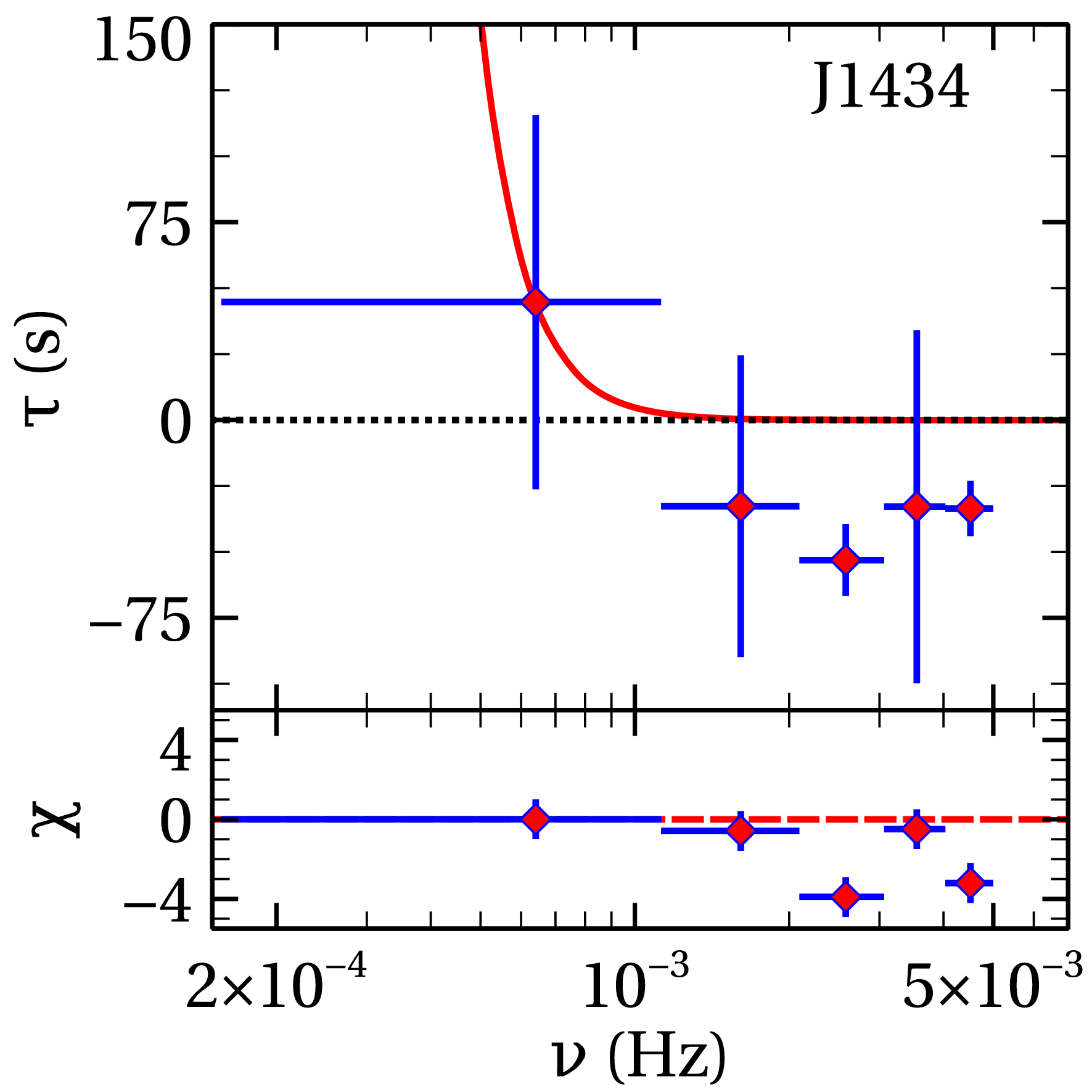}
\includegraphics[scale=0.32,angle=-0]{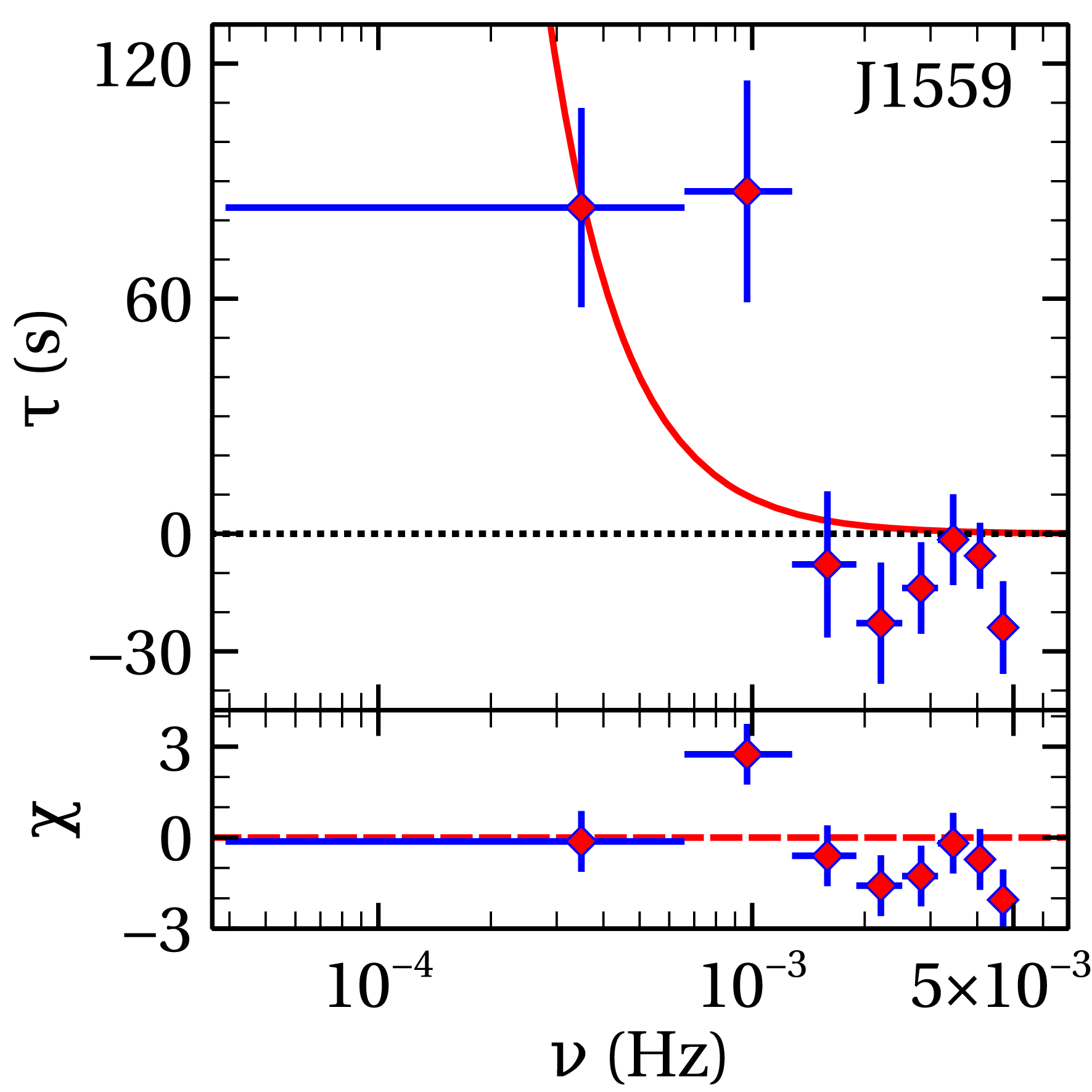}
\includegraphics[scale=0.32,angle=-0]{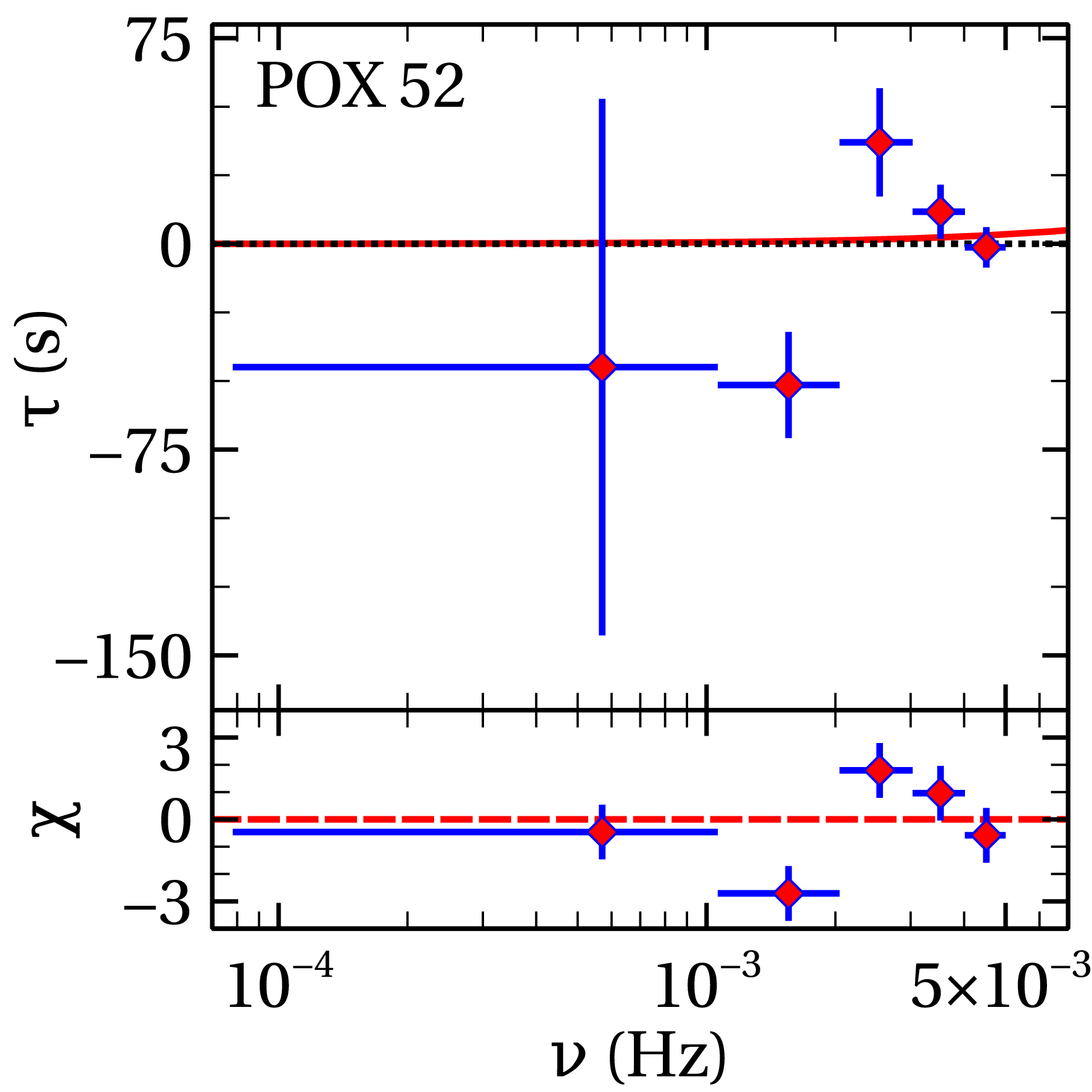}
\caption{Lag-frequency spectra fitted by a monotonically declining hard lag model, $\tau_{h}=k(\nu/\nu_{0})^{-\alpha}$. The red, solid line shows the fitted hard lag model. The residuals shown in the lower panels demonstrate the existence of negative soft lags in the spectra. However, the Monte Carlo simulations suggest that the observed soft lags are only marginally significant for J0942 and POX52 with confidence $<90\%$.}
\end{center}
\label{fig_extra2}
\end{figure*}

\bsp	
\label{lastpage}
\end{document}